\documentclass[aps,prd,
floatfix,nofootinbib,groupedaddress,showpacs]{revtex4-1}

\usepackage{amsmath}
\usepackage{amssymb}
\usepackage{graphicx}

\usepackage{latexsym}
\usepackage{array}
\usepackage{verbatim}

\usepackage{rotating}
\usepackage{color}

\usepackage[utf8]{inputenc}
\usepackage{threeparttable}

\def\lsim{\mathrel{\raise.3ex\hbox{$<$\kern-.75em\lower1ex\hbox{$\sim$}}}}
\def\gsim{\mathrel{\raise.3ex\hbox{$>$\kern-.75em\lower1ex\hbox{$\sim$}}}}

\newcommand{\be}{\begin{equation}}
\newcommand{\ee}{\end{equation}}
\def\ddel{\!\!\mathrel{\raise1.7ex\hbox{$\leftrightarrow$\kern-.85em
\lower1.7ex\hbox{$\partial$}}}\!}
\allowdisplaybreaks

\newlength{\absize}
\newcommand{\half}{{\textstyle\frac{1}{2}}}

\def\lsim{\mathrel{\rlap{\raise 2.5pt \hbox{$<$}}\lower 2.5pt
\hbox{$\sim$}}}

\newcommand{\Tr}{{\rm Tr}}

\allowdisplaybreaks

\begin{document}

%------------------------------------
\title{$S_3$-inspired three-Higgs-doublet models: A class with a complex vacuum }
\author{A. Kun\v cinas}
\email{Anton.Kuncinas@protonmail.com}
\affiliation{Department of Physics and Technology, University of Bergen,
Postboks 7803, N-5020  Bergen, Norway}
\author{O. M. Ogreid}
\email{omo@hvl.no}
\affiliation{Western Norway University of Applied Sciences, Postboks 7030, N-5020 Bergen, 
Norway}
\author{P. Osland}
\email{Per.Osland@uib.no}
\affiliation{Department of Physics and
Technology, University of Bergen, Postboks 7803, N-5020 Bergen,
Norway}
\author{M. N. Rebelo}
\email{rebelo@tecnico.ulisboa.pt. On leave of absence from CFTP/IST, U. Lisboa}
\affiliation{Centro de F\'isica Te\'orica de Part\'iculas -- CFTP and Dept de F\' \i sica
Instituto Superior T\'ecnico -- IST, Universidade de Lisboa, Av. Rovisco Pais,
P-1049-001 Lisboa, Portugal}
\affiliation{CERN, Theoretical Physics Department, CH-1211 Geneva 23, Switzerland}

\date{\today}

\begin{abstract}

In this paper we analyse in detail an $S_3$-symmetric three-Higgs-doublet
model with a specific vacuum configuration. This analysis allows us to illustrate 
important features of models with several Higgs doublets, such as 
the possibility of having spontaneous CP violation. We start with a real potential
and pick a particularly interesting complex vacuum configuration, which 
does not violate CP before adding soft breaking terms to the potential. We study the r\^ ole played by 
different soft symmetry breaking terms. These are essential for our choice of vacuum 
in order to remove unwanted massless scalars which arise from the spontaneous breaking of an accidental continuous symmetry. We list scalar sector and scalar-gauge sector-couplings for the particular case we consider in detail in this work.
Results presented in this paper will be useful for model building, in particular for implementations of models with $S_3$ symmetry and spontaneous CP violation, extensions of the fermionic sector with realistic Yukawa couplings and for Dark Matter studies.
\end{abstract}

\maketitle

%%%%%%%%%%%%%%%%%%%%%%%%%%%%%%%%%%%%%%%%%%%%%
\section{Introduction}
%%%%%%%%%%%%%%%%%%%%%%%%%%%%%%%%%%%%%%%%%%%%%

In the Standard Model there is only one Higgs doublet, leading, through the
Higgs mechanism \cite{Higgs:1964ia,Higgs:1964pj,Englert:1964et,Guralnik:1964eu}, to the existence of one Higgs boson. Such a particle was discovered in 2012 at the LHC \cite{Aad:2012tfa,Chatrchyan:2012xdj}. The question remains of whether or not
there are additional Higgs bosons in Nature. Multi-Higgs extensions of the Standard Model (SM) are
very well motivated. In particular a lot of work has been done in the context of models
with two Higgs doublets, for reviews see \cite{Gunion:1989we,Branco:2011iw}
as well as in models with three or more  Higgs doublets
\cite{Ivanov:2017dad}. As the complexity of the Higgs sector grows the number of free parameters increases \cite{Olaussen:2010aq}.

Symmetries play an important role in controlling the
number of free parameters, therefore increasing the predictability of such extensions.
In these scenarios some of the problems of the SM, such as the need for new sources of CP violation,
can be addressed. These models have a rich phenomenology and will be tested at the LHC and future colliders. An important feature of multi-Higgs extensions of the SM is the possibility of
having spontaneous CP violation. It was shown by T. D. Lee \cite{Lee:1973iz} that models with two Higgs doublets can violate CP spontaneously. Imposing additional symmetries may eliminate the possibility of having
spontaneous CP violation. On the other hand, continuous symmetries broken by vacuum expectation
values (vevs) lead to the existence of massless scalars 
\cite{Nambu:1960tm,Goldstone:1961eq,Goldstone:1962es}. These are ruled out by 
experiment. There are also
strong experimental constraints which have to be taken into consideration when extending the
Higgs sector of the SM \cite{Tanabashi:2018oca}.

In this paper, we revisit the different vacuum solutions for the $S_3$-symmetric potential with three Higgs doublets which were studied previously in Ref.~\cite{Emmanuel-Costa:2016vej}. The $S_3$-symmetric scalar sector with three Higgs doublets was studied in the past by several authors, starting in 1977 by Pakvasa and Sugawara \cite{Pakvasa:1977in} who worked with irreducible representations consisting of a doublet and singlet of $S_3$ and also analyzed couplings to fermions. Derman and Tsao \cite{Derman:1978rx,Derman:1979nf} shortly afterwards discussed several properties of these models in terms of the defining representation of $S_3$.
In Ref.~\cite{Emmanuel-Costa:2016vej}, special attention was paid to the possibility of having spontaneous CP violation. However, it was also pointed out that several potentially interesting vacua led to the existence of massless scalars \cite{Emmanuel-Costa:2016vej,Kuncinas}. These massless states are due to the spontaneous breaking of accidental continuous symmetries resulting from constraints imposed on the region of parameters arising from the minimisation conditions, with a few rare exceptions that will be pointed out in the paper.
The identification of such symmetries is of great relevance and has been dealt with in the context of general three-Higgs doublet  models (3HDMs) by several authors \cite{deMedeirosVarzielas:2019rrp,Darvishi:2019dbh} (see also Ref.~\cite{Ishimori:2012zz}).

Here, we work with the irreducible representations, doublet and singlet and
we introduce soft breaking terms (terms bilinear in the fields) in the potential, which break the symmetries leading to massless scalars. First, we consider all possible forms for the soft breaking terms and we separate the vacuum solutions according to whether they are real or complex. Next, we classify them according to the number of zero vacuum expectation values and their positions. The rest of our analysis centres on the study of a specific vacuum solution of the unbroken $S_3$ symmetry which suffers from unwanted massless bosons. This we call the C-III-c solution.

Before introducing soft breaking terms, the C-III-c vacuum displays very curious properties. It is the only vacuum allowing for a non-trivial phase, which is not determined by any of the parameters of the potential and remains free. However, there is a term in the potential which is sensitive to this phase, denoted $\lambda_7$ below. This term accounts for a coupling between two fields from the $S_3$ doublet and two fields from the $S_3$ singlet. Since the C-III-c vacuum and its generalisations when soft breaking terms are included, have a vanishing singlet vev, this dependence on the phase will enter the mass-squared matrix for the singlet states. This phase will also enter in trilinear couplings.
When soft breaking terms are introduced, this phase will also have an impact on other parts of the potential.

This paper is organised as follows. In section~\ref{sect:framework} we explain the general framework together with a brief discussion of some of the vacuum solutions obtained in \cite{Emmanuel-Costa:2016vej}. In section~\ref{sect:goldstone} we discuss the origin of the massless states and in section~\ref{sect:C-III-c} we review some properties of the C-III-c model.
In section~\ref{sect:soft} we list all possible forms for the vacua in terms of zero vevs and we discuss the effects of each allowed soft breaking term. Next, in section~\ref{sect:C-III-c-soft} we focus our attention on the discussion of a class of $S_3$-inspired 3HDMs with a complex vacuum characterised by having zero vev for the $S_3$ singlet and with the two other vevs being arbitrary complex. Previously, we called this vacuum C-III-c before introducing soft breaking terms. In Ref.~\cite{Emmanuel-Costa:2016vej}  we had shown that without soft breaking terms this vacuum did not violate CP spontaneously, despite being complex. We now show what effect the introduction of the different soft breaking terms can have on the CP properties of this complex vacuum. Finally, in the last section, we present our summary.
Results presented in this paper will be important for model building, in particular for implementations of models with $S_3$ symmetry and spontaneous CP violation, extensions of the fermionic sector with realistic Yukawa couplings and for Dark Matter studies.

%%%%%%%%%%%%%%%%%%%%%%%%%%%%%%%%%%%%%%%%%%%%%
\section{Framework} \label{sect:framework}
\setcounter{equation}{0}
%%%%%%%%%%%%%%%%%%%%%%%%%%%%%%%%%%%%%%%%%%%%%
The $S_3$ symmetry is a symmetry for the permutation of three objects, in this case three Higgs doublet fields, $\phi_1$, $\phi_2$ and $\phi_3$.

The scalar potential expressed in terms of the $S_3$ irreducible representation singlet
and doublet fields, respectively  ($h_S$) and ($h_1, h_2$), can be written 
as
\begin{equation}\label{Eq:pot-24}
V=V_2+V_4,
\end{equation}
with
\cite{Kubo:2004ps,Teshima:2012cg,Das:2014fea}:
\begin{subequations}
\begin{align}
V_2&=\mu_0^2 h_S^\dagger h_S +\mu_1^2(h_1^\dagger h_1 + h_2^\dagger h_2), \\
V_4&=
\lambda_1(h_1^\dagger h_1 + h_2^\dagger h_2)^2 
+\lambda_2(h_1^\dagger h_2 - h_2^\dagger h_1)^2
+\lambda_3[(h_1^\dagger h_1 - h_2^\dagger h_2)^2+(h_1^\dagger h_2 + h_2^\dagger h_1)^2]
\nonumber \\
&+ \lambda_4[(h_S^\dagger h_1)(h_1^\dagger h_2+h_2^\dagger h_1)
+(h_S^\dagger h_2)(h_1^\dagger h_1-h_2^\dagger h_2)+ \hbox {h.c.}] 
+\lambda_5(h_S^\dagger h_S)(h_1^\dagger h_1 + h_2^\dagger h_2) \nonumber \\
&+\lambda_6[(h_S^\dagger h_1)(h_1^\dagger h_S)+(h_S^\dagger h_2)(h_2^\dagger h_S)] 
+\lambda_7[(h_S^\dagger h_1)(h_S^\dagger h_1) + (h_S^\dagger h_2)(h_S^\dagger h_2) 
+\hbox {h.c.}]
\nonumber \\
&+\lambda_8(h_S^\dagger h_S)^2.
\label{Eq:V-DasDey-quartic}
\end{align}
\end{subequations}
where we are taking all coefficients to be real. Therefore, there is no explicit CP violation.

For the SU(2) doublets we use the notation
\begin{equation} \label{Eq:hi_hS}
h_i=\left(
\begin{array}{c}h_i^+\\ (w_i+\eta_i+i\chi_i)/\sqrt{2}
\end{array}\right), \quad i=1,2, \quad
h_S=\left(
\begin{array}{c}h_S^+\\ (w_S+ \eta_S+i \chi_S)/\sqrt{2}
\end{array}\right).
\end{equation}
In some cases, it is also convenient to extract an overall phase.

The irreducible representations can be related to the defining set of Higgs doublets $\phi_1$, $\phi_2$ and $\phi_3$, by:
\begin{equation} \label{Eq:reps}
\left( \begin{array}{c}
h_1\\
h_2\\
h_S \\
\end{array}  \right) = \left( \begin{array}{ccc} 
\frac{1}{\sqrt{2}} & \frac{-1}{\sqrt{2}} & 0 \\ 
\frac{1}{\sqrt{6}} & \frac{1}{\sqrt{6}} & \frac{-2}{\sqrt{6}} \\
\frac{1}{\sqrt{3}} & \frac{1}{\sqrt{3}} & \frac{1}{\sqrt{3}} \\
\end{array} \right) 
 \left( \begin{array}{c}
\phi_1 \\
\phi_2 \\
\phi_3 \\
\end{array}  \right).
\end{equation}

Early papers studying this potential were written in terms of the defining representation \cite{Derman:1978rx,Derman:1979nf}. Notice that Eq.~(\ref{Eq:reps}) chooses a particular direction in the space of doublets of the defining representation. There is no freedom in the direction of $h_S$, where they all play an equal role. However, for $h_1$, any permutation of $(1,-1,0)$ is equally valid, implying the corresponding changes in the definition of $h_2$.
There is no physics content in any of these arbitrary choices, but they translate differently in terms of consistency conditions when going from the defining representation to the irreducible representation. This should be clear from the discussion in section~4 of Ref.~\cite{Emmanuel-Costa:2016vej}. Further comments will be presented below.

A full classification of all possible vacua, together with the necessary constraints on the parameters of the potential coming from the minimisation conditions, was given in Ref.~\cite{Emmanuel-Costa:2016vej}.
In appendix~\ref{app:A} we present a summary of some of their properties.
There are 11 vacua with all vevs real (table~\ref{Table:real}), which were denoted R followed by a further specification, and 17 vacua with at least one complex vev (table~\ref{Table:complex}), for which the letter C is used. Each vacuum is labeled by R or C followed by a Roman number and possibly another alphabetic index. The Roman number indicates how many constraints on the parameters are required by the minimisation conditions. The additional alphabetic index is used to distinguish different (real or complex) vacua with the same number of constraints.

It is instructive to summarise some properties of the possible vacua of the $S_3$-symmetric potential. 
Table~\ref{Table:real} in appendix~\ref{app:A} lists all possible real vacuum solutions, with $\lambda_a$ defined by:
\begin{subequations} \label{Eq:lambda_ab}
\begin{align}
\lambda_a&=\lambda_5+\lambda_6+2\lambda_7, \\
\lambda_b&=\lambda_5+\lambda_6-2\lambda_7,
\end{align}
\end{subequations}
(the quantity $\lambda_b$ is be used in the discussion of the complex vacua).

All real vacuum solutions other than R-0 and R-I-1 violate the $S_3$ symmetry spontaneously. For these solutions, the residual symmetries were discussed by Derman and Tsao \cite{Derman:1979nf}.

The constraints given in table~\ref{Table:real} come from the stationary-point conditions\footnote{While all consistent solutions correspond to local minima of the potential, it is not yet known whether they all correspond to global minima. This issue has been addressed in the context of 2HDMs \cite{Ferreira:2004yd,Barroso:2005sm}, whereas for 3HDMs \cite{Ivanov:2010wz} it is not yet fully analysed. Such results would give an important contribution to the understanding of 3HDMs.}
\begin{subequations} \label{Eq:DD-mu0-mu1}
\begin{align} \label{Eq:mu_0_sq}
2\mu_0^2w_S
+ \lambda_4(3w_1^2-w_2^2)w_2 +\lambda_a(w_1^2+w_2^2)w_S +2\lambda_8w_S^3
&=0, \\
\left[2\mu_1^2+
2(\lambda_1+\lambda_3) (w_1^2+w_2^2)+6\lambda_4w_2w_S +\lambda_a w_S^2
\right]w_1&=0, \label{Eq:DD-mu_1-a}\\
2\mu_1^2w_2+
2(\lambda_1+\lambda_3) (w_1^2+w_2^2)w_2
+3\lambda_4(w_1^2-w_2^2)w_S +\lambda_a w_2w_S^2 
&=0, \label{Eq:DD-mu_1-b} 
\end{align}
\end{subequations}
which were discussed in Ref.~\cite{Emmanuel-Costa:2016vej}. We see that in order to classify all possible solutions to these equations, one must consider all configuration of the vacuum parameters $w_i$, in particular all possible situations where two of the vevs vanish as well as all the possible situations where only one vev vanishes, and finally the situation where none of the vevs vanishes (R-III). For each configuration of vacuum parameters, we determine the resulting constraints on the parameters of the potential, in each case solving for $\mu_0^2$ and  $\mu_1^2$ (if possible).
In particular, since both Eqs.~(\ref{Eq:DD-mu_1-a}) and (\ref{Eq:DD-mu_1-b}) contain $\mu_1^2$, in order to solve them simultaneously we will get an additional constraint either on the parameters of the vevs or on the parameters of the potential.
For all vacua other than R-III, this translates into restrictions on the parameter space as well as restrictions on the allowed vevs. 

For $w_1=0$, the derivative given by Eq.~(\ref{Eq:DD-mu_1-a}) is automatically zero and there is no clash. Otherwise, the terms proportional to $\lambda_4$ in Eqs.~(\ref{Eq:DD-mu_1-a}) and (\ref{Eq:DD-mu_1-b}) must be restricted, requiring 
\begin{equation}
\lambda_4(3w_2^2-w_1^2)w_S=0.
\label{Eq:lambda4=0,DD}
\end{equation}
This can be achieved by having $\lambda_4=0$ or $w_1=\pm\sqrt3 w_2$ or $w_S=0$.
For $w_S=0$, an additional condition arises from Eq.~(\ref{Eq:mu_0_sq}),
\begin{equation} \label{Eq:cond-2}
\lambda_4 (3w_1^2-w_2^2)w_2=0,
\end{equation}
which implies that either $\lambda_4=0$ or $w_2=\pm\sqrt3 w_1$ or else $w_2=0$.

The cases R-I-2 satisfy both Eqs.~(\ref{Eq:lambda4=0,DD}) and (\ref{Eq:cond-2}) without requiring $\lambda_4=0$. In terms of the defining representation they are obviously equivalent. In terms of the irreducible representation they obey Eq.~(\ref{Eq:cond-2}) by having $w_2=0$ in case a, $w_2=\sqrt3 w_1$ in case~b and $w_2=-\sqrt3 w_1$ in case~c. This difference is the result of having chosen a particular direction for $h_1$, as pointed out above. In this case, the residual symmetry is $S_2$. Another interesting set of cases with residual symmetry $S_2$ are the R-II-1 vacua, which are also equivalent in terms of the defining representation and obey the consistency conditions with $w_1=0$ in case~a, $w_1=-\sqrt3 w_2$ in case~b and $w_1=\sqrt3 w_2$ in case~c. The vacua R-II-2, R-II-3 and R-III all require $\lambda_4=0$.

A particularly interesting vacuum is the one identified as case C-III-c,
\begin{equation}
(w_1,w_2,w_S) = (\hat w_1e^{i\sigma_1},\hat w_2e^{i\sigma_2},0),
\end{equation}
with $\hat w_1$ and $\hat w_2$ real and positive.
The three stationary-point constraints are:
\begin{subequations} \label{Eq:C-III-c-conditions}
\begin{align}
&\mu_1^2=-(\lambda_1+\lambda_3)(\hat{w}_1^2+\hat{w}_2^2),\\
&\lambda_2+\lambda_3=0, \\
&\lambda_4=0. \label{Eq:lambda40}
\end{align}
\end{subequations}

In this solution, $\lambda_4$ is required to be zero. It should be pointed out that $\lambda_4$ plays a very important role in this potential.
Whenever $\lambda_4\neq0$, the only independent Higgs-family symmetries of the potential are the global U(1) symmetry and the symmetry under which $h_1$ changes sign. The combination of these two symmetries also yields a potential that is symmetric under the simultaneous change of sign of both $h_2$ and $h_S$.
On the other hand, whenever $\lambda_4=0$, the potential acquires an additional (continuous) independent O(2) symmetry between the fields $h_1$ and $h_2$. Combining this symmetry with the two symmetries that were already present without a vanishing  $\lambda_4$, one finds that now, with $\lambda_4=0$, the potential is also symmetric under the independent sign changes of either $h_2$ or $h_S$.

This O(2)  symmetry allows for the rotation between $h_1$ and $h_2$ in such a way that, together with an appropriate overall rephasing, this vacuum can be transformed into \cite{Ogreid:2017alh}
\begin{equation} \label{Eq:vac-sigmahalf}
(w_1,w_2,w_S) = (\hat we^{i\sigma/2},\hat we^{-i\sigma/2},0),
\end{equation}
so that it can be easily shown \cite{Branco:1983tn,Ogreid:2017alh} that it does not violate CP spontaneously.

The continuous O(2) symmetry is spontaneously broken by the C-III-c vacuum, giving rise to a massless neutral scalar field.  However, we have one more massless neutral scalar for which there is no explanation in terms of a spontaneously broken continuous symmetry. We shall comment on its origin below.
In all, for the C-III-c case, we have {\it two} massless neutral scalars
which need to be removed in order to construct a physical model. One way to achieve this is by adding one or more terms that break the $S_3$ symmetry softly.

%%%%%%%%%%%%%%%%%%%%%%%%%%%%%%%%%%%%%%%%%%%%%
\section{Goldstone bosons} \label{sect:goldstone}
\setcounter{equation}{0}
%%%%%%%%%%%%%%%%%%%%%%%%%%%%%%%%%%%%%%%%%%%%%
Several of the possible vacuum solutions of the $S_3$-symmetric potential have massless scalars. These result from the spontaneous breakdown of accidental continuous symmetries that arise when we impose the constraints required for these solutions. In tables~\ref{Table:S3-real-vacua} and \ref{Table:S3-complex-vacua} we list the number of massless scalars for each case, together with whether or not $\lambda_4$ is required to be zero. 

For $\lambda_4=0$ the potential acquires an additional O(2) symmetry between the two members of the $S_3$ doublet. 
When this symmetry is broken by the vacuum, one massless scalar state appears. In some cases, $\lambda_7$ is also required to be zero, together with $\lambda_4=0$, and the potential acquires an additional U(1) symmetry which we denote by $\text{U(1)}_{h_s}$. This corresponds to the freedom of rephasing $h_S$ independently from $h_1$ and $h_2$. Once again, an additional massless scalar state appears when this symmetry is spontaneously broken.
In the C-III-c case, the condition $\lambda_4=0$ is accompanied by
$\lambda_2+\lambda_3=0$. This last condition does not increase the symmetry, therefore it may not be exactly preserved at all energy scales.\footnote{We thank Pedro Ferreira for illustrating this feature in a private discussion.}
However, there are two massless states in the C-III-c case, as discussed in section~\ref{sect:C-III-c}.
Note that there is no vacuum which requires $\lambda_2+\lambda_3=0$ or
$\lambda_7=0$ without also having $\lambda_4=0$. 

%%%%%%%%%%%%%%%%%%%%%%%%%%%%%%%%%%%%%%%%%%%%%
\begin{table}[htb]
\caption{Real vacua, for the unbroken $S_3$ case, with massless states and degeneracies indicated. The first entry in the parenthesis refers to the charged sector, the second one to the neutral sector.}
\label{Table:S3-real-vacua}
\begin{center}
\begin{tabular}{|c|c|c|c|c|c|c|c|}
\hline\hline
Vacuum & name & $\lambda_4$ & symmetry &\# massless states & degeneracies  \\
\hline
\hline
$R_{00x}$ & R-I-1 & $\surd$ & & none & (1,2) \\
\hline
$R_{0x0}$ & R-II-2 & 0  & O(2) & (none,1) &none \\
\hline
$R_{x00}$ & R-I-2a & $\surd$  &  & none & none \\
\hline
$R_{0xy}$ & R-II-1a & $\surd$ &  & none & none \\
\hline
$R_{x0y}$  & & 0 & O(2) & (none,1) & none  \\
\hline
$R_{xy0}$ & R-I-2b,2c & $\surd$ & & none & none  \\
\hline
$R_{xy0}$ & R-II-3 & 0 & O(2) & (none,1) & none  \\
\hline
$R_{xyz}$ & R-II-1b,1c & $\surd$ & & none & none  \\
\hline
$R_{xyz}$ & R-III & 0 & O(2) & (none,1) & none  \\
\hline
\hline
\end{tabular}
\end{center}
\end{table}
%%%%%%%%%%%%%%%%%%%%%%%%%%%%%%%%%%%%%%%%%%%%%

In the C-V case, all of these are required to be zero, in this case we can independently rephase any of 
the doublets $h_1$, $h_2$ and $h_S$ and therefore there are three  U(1) symmetries which we denote by 
$\text{U(1)}_{h_1}$, $\text{U(1)}_{h_2}$ and $\text{U(1)}_{h_S}$. The spontaneous breakdown of the resulting symmetry  
$\text{O(2)}\otimes\text{U(1)}_{h_1}\otimes\text{U(1)}_{h_2}\otimes\text{U(1)}_{h_S}$ is responsible for the three 
massless states that appear in addition to the would-be Goldstone boson.  
Note that with an overall phase rotation of the three doublets 
one can always reduce an independent rephasing of the three doublets to an independent rephasing of any 
two of them. Additional U(1) symmetries only arise in cases with $\lambda_4 = 0$.

%%%%%%%%%%%%%%%%%%%%%%%%%%%%%%%%%%%%%%%%%%%%%
\begin{table}[htb]
	\caption{Complex vacua, for the unbroken $S_3$ case, with massless states and degeneracies indicated,
	not taking into account would-be Goldstone bosons.
	The first entry in the parenthesis refers to the charged sector, the second one to the neutral sector.
	Degeneracies only refer to massive pairs.
		In the footnotes below, $L$  indicates that a linear expression in its arguments vanishes.}
	\label{Table:S3-complex-vacua}
	\begin{center}
		\begin{threeparttable}
			\begin{tabular}{|c|c|c|c|c|c|c|c|}
				\hline\hline
				Vacuum & name & \phantom{m}$\lambda_4$\phantom{m} & symmetry &\# massless states & degeneracies  \\
				\hline
				\hline
				$C_{0xy}$ & C-III-a & $\surd$ &  & none & none \\
				\hline
				$C_{x0y}$ & C-III-b & 0 &  O(2) & (none,1) & none \\
				\hline
				$C_{x0y}$ & C-IV-a & 0\tnote{$\alpha$} &  $\text{O(2)}\otimes\text{U(1)}_{h_S}$ & (none,2) 
				& none \\
				\hline
				$C_{xy0}$ & C-I-a &  $\surd$ & & none & (none,2) \\
				\hline
				$C_{xy0}$ & C-III-c &  0\tnote{$\beta$} & O(2) & (none,2) & none \\
				\hline\hline
				$C_{xyz}$ & C-III-d,e & $\surd$ &  & none & none \\
				\hline
				$C_{xyz}$ & C-III-f,g & 0 & O(2) & (none,1) & none \\
				\hline
				$C_{xyz}$ & C-III-h,i & $\surd$ &  & none & none \\
				\hline
				$C_{xyz}$ & C-IV-b & 0 & O(2) & (none,1) & none \\
				\hline
				$C_{xyz}$ & C-IV-c & \tnote{$\gamma$} & -  & (none,1) & none \\
				\hline
				$C_{xyz}$ & C-IV-d & 0\tnote{$\alpha$} & $\text{O(2)}\otimes\text{U(1)}_{h_S}$ & (none,2) & none \\
				\hline
				$C_{xyz}$ & C-IV-e & 0 & O(2) & (none,1) & none \\
				\hline
				$C_{xyz}$ & C-IV-f & \tnote{$\gamma$} & - & (none,1) & none \\
				\hline
				$C_{xyz}$ & C-V & 0\tnote{$\alpha$,$\beta$} & $\text{O(2)}\otimes\text{U(1)}_{h_1}\otimes\text{U(1)}_{h_2}\otimes\text{U(1)}_{h_S}$  & (none,3) & none \\
				\hline
				\hline
			\end{tabular}
			\begin{tablenotes}\footnotesize
				\item[{$\alpha$}] Also $\lambda_7=0.$
				\hspace{20mm} ${}^\beta$ Also $\lambda_2+\lambda_3=0.$
				\hspace{20mm} ${}^\gamma$ $L(\lambda_2+\lambda_3,\lambda_4)$, $L(\lambda_2+\lambda_3,\lambda_7)$.
			\end{tablenotes}
		\end{threeparttable}
	\end{center}
\end{table}
%%%%%%%%%%%%%%%%%%%%%%%%%%%%%%%%%%%%%%%%%%%%%

Tables~\ref{Table:S3-real-vacua} and \ref{Table:S3-complex-vacua} illustrate the fact that there are several vacua of the same generic form, when expressed for instance in terms of vevs of the irreducible representations,  as specified in the first column, which have different physical implications. This is due to the fact that the same generic solution can be obtained for different regions of parameters. Different regions may also lead to different accidental symmetries.

%%%%%%%%%%%%%%%%%%%%%%%%%%%%%%%%%%%%%%%%%%%%%
\section{The C-III-c model without soft breaking terms} \label{sect:C-III-c}
\setcounter{equation}{0}
%%%%%%%%%%%%%%%%%%%%%%%%%%%%%%%%%%%%%%%%%%%%%

The C-III-c model (based on the ``$\sigma$ vacuum'') has some peculiar properties. As mentioned above, it has two massless states in the neutral sector (apart from the would-be Goldstone boson). Removing them is the main purpose of introducing soft $S_3$-breaking terms. This will be done in the next section. As pointed out before, in the C-III-c case the condition $\lambda_4=0$ is accompanied by $\lambda_2+\lambda_3=0$. This last condition does not increase the symmetry.
If in addition we were to have $\lambda_7=0$ then new continuous symmetries of the potential would exist just like in case C-V. One may wonder then why, in this case, one has two massless scalars rather then only one. The reason has to do with the fact that there is no $\lambda_7$ term in the mass terms of the scalar fields coming from the $S_3$ doublet. This fact is accidental, it results from this vacuum configuration having $\hat w_S=0$. The mass terms for these fields of the $S_3$ doublet mimic the existence of a larger symmetry under which $h_1$ and $h_2$ may be independently rephased, since we have $\lambda_2 + \lambda_3 =0$ and $\lambda_4=0$. The vacuum is not invariant under this rephasing of $h_1$ and $h_2$. However, rephasing of $h_S$ independently from the other doublets leaves the vacuum invariant. Invariance under an overall rephasing of the three scalars implies that only one of these former two U(1) transformations is independent.

The technique proposed in Ref.~\cite{deMedeirosVarzielas:2019rrp} is a useful tool to search for symmetries, in the context of three-Higgs-doublet models, that are not explicit. That method applied to this case confirms the non-existence of an additional continuous symmetry. 
In subsections \ref{C-III-c-masses} and \ref{C-III-c-trilinear} we shall be confronted with the fact that the mass splitting of the neutral scalars in the $S_3$ singlet sector and some trilinear couplings will depend on the relative phase of the two vevs, $\sigma$, which seems to be unrelated to the coefficients of the potential. This apparent paradox is addressed in subsection~\ref{Eq:subsect-sigma}.

Examples of the connection between symmetries and mass degeneracies in two- and three-Higgs-doublet models with vanishing vevs can be found in Ref.~\cite{Haber:2018iwr}.

%%%%%%%%%%%%%%%%%%%%%%%%%%%%%%%%%%%%%%%%%%%%%%%%%%%%%%%
\subsection{Masses}
\label{C-III-c-masses}
%%%%%%%%%%%%%%%%%%%%%%%%%%%%%%%%%%%%%%%%%%%%%%%%%%%%%%%

Since $\lambda_4=0$ and $\hat w_S=0$, the $S_3$ doublet and the $S_3$ singlet do not mix in the mass terms. In the charged sector, we have
\begin{align}
m^2_{H^\pm}&=2\lambda_2 v^2, \\
m^2_{S^\pm}&=\mu_0^2+\half\lambda_5v^2,
\end{align}
where $v^2=\hat w_1^2+\hat w_2^2$ and $H^\pm$ and $S^\pm$ refer to the charged states of the doublet and singlet sector, respectively.

In the neutral sector of the $S_3$ doublet, there is only one massive (CP-even) state, 
\begin{equation}
m^2_{h}=2(\lambda_1-\lambda_2)v^2,
\end{equation}
which would have to be identified with the SM-like Higgs, since it appears in the doublet where the would-be Goldstone bosons are. There is no further mixing with the other fields.
The $S_3$ singlet sector has two massive states ($S_1$ and $S_2$),
\begin{subequations} \label{Eq:mass-splitting-singlet}
\begin{align}
m^2_{S_1}&=\mu_0^2+\half(\lambda_5+\lambda_6)v^2-\lambda_7\cos\sigma v^2, \\
m^2_{S_2}&=\mu_0^2+\half(\lambda_5+\lambda_6)v^2+\lambda_7\cos\sigma v^2.
\end{align}
\end{subequations}
Thus, the phase $\sigma$, which is left undetermined by the potential, is related to the mass splitting of the neutral scalars in the $S_3$ singlet sector.

%%%%%%%%%%%%%%%%%%%%%%%%%%%%%%%%%%%%%%%%%%%%%%%%%%%%%%%
\subsection{Gauge couplings}
%%%%%%%%%%%%%%%%%%%%%%%%%%%%%%%%%%%%%%%%%%%%%%%%%%%%%%%

For the different couplings, we should define the degenerate fields carefully. Thus, rather than adopting the decomposition (\ref{Eq:hi_hS}), we take
\begin{equation}
h_1=e^{i\sigma/2}\left(
\begin{array}{c}h_1^+\\ (\hat w+\eta_1+i\chi_1)/\sqrt{2}
\end{array}\right), \quad 
h_2=e^{-i\sigma/2}\left(
\begin{array}{c}h_2^+\\ (\hat w+\eta_2+i\chi_2)/\sqrt{2}
\end{array}\right), \label{def}
\end{equation}
and
\begin{equation}
h_S=\left(
\begin{array}{c}S^+\\ (S_1+i S_2)/\sqrt{2}
\end{array}\right),
\end{equation}
with $\hat w^2=v^2/2$.
Since the $S_3$ singlet has a vanishing vev, it is straightforward to transform to the Higgs basis \cite{Donoghue:1978cj,Georgi:1978ri}. A convenient choice is to leave the singlet fields as they are, and take
\begin{alignat}{2}
h_1^\pm&=(G^\pm-H^\pm)/\sqrt2, &\quad h_2^\pm&=(G^\pm+H^\pm)/\sqrt2, \\
\eta_1&=(h-H)/\sqrt2, &\quad \eta_2&=(h+H)/\sqrt2, \\
\chi_1&=(G^0-A)/\sqrt2, &\quad \chi_2&=(G^0+A)/\sqrt2.
\end{alignat}
This choice fixes the definitions of the degenerate, massless bosons $H$ and $A$ (any orthogonal basis would be equally good).

The covariant derivatives induce gauge couplings, those linear in the gauge fields are
\begin{align}
{\cal L}_{VHH}
&=ieA^\mu[(H^+\ddel_\mu H^-)+(S^+\ddel_\mu S^-)] \nonumber \\
&-\frac{g}{2\cos\theta_W}Z^\mu\{(H\ddel_\mu A)+(S_1\ddel_\mu S_2) 
 -i\cos2\theta_W[(H^+\ddel_\mu H^-)+(S^+\ddel_\mu S^-)]\} \nonumber \\
&+\frac{ig}{2}\{W^\mu[(H\ddel_\mu H^-)+i(A\ddel_\mu H^-)
+(S_1\ddel_\mu S^-)+i(S_2\ddel_\mu S^-)]
-\text{h.c.}\},
\end{align}
where $\theta_W$ is the weak mixing angle.
Furthermore, $H$ and $A$ denote the CP-even and odd massless states.
Next, the terms bilinear in gauge fields are
\begin{align} \label{Eq:L-VVH}
{\cal L}_{VVH}
&=\frac{g^2v}{4\cos^2\theta_W}Z_\mu Z^\mu h
+\frac{g^2 v}{2}W^\dagger_\mu W^\mu h,
\end{align}
and
\begin{align}
{\cal L}_{VVHH}
&=\left[\left(eA +\frac{g\cos2\theta_W}{2\cos\theta_W} Z\right)_\mu
\left(eA +\frac{g\cos2\theta_W}{2\cos\theta_W} Z\right)^\mu
+\frac{g^2}{2}W^\dagger_\mu W^\mu\right]
(H^-H^++S^-S^+) \nonumber \\
&+\left[\frac{g^2}{8\cos^2\theta_W}Z_\mu Z^\mu
+\frac{g^2}{4}W^\dagger_\mu W^\mu\right]
(h^2+H^2+A^2+S_1^2+S_2^2) \nonumber \\
&+\frac{eg}{2}\{A_\mu W^\mu[H^-(H+iA)+S^-(S_1+iS_2)] 
+\text{h.c.}\} \nonumber \\
&-\frac{g^2\sin^2\theta_W}{2\cos\theta_W}\{Z_\mu W^\mu
[H^-(H+iA)+S^-(S_1+iS_2)] +\text{h.c.}\}.
\end{align}
Note that ${\cal L}_{VVH}$ contains no term linear in the singlet fields since it has a vanishing vev. 

%%%%%%%%%%%%%%%%%%%%%%%%%%%%%%%%%%%%%%%%%%%%%%%%%%%%%%%
\subsection{Trilinear couplings}
\label{C-III-c-trilinear}
%%%%%%%%%%%%%%%%%%%%%%%%%%%%%%%%%%%%%%%%%%%%%%%%%%%%%%%
The non-zero trilinear couplings are (as coefficients of the potential):
\begin{subequations} \label{Eq:couplings-trilin}
\begin{alignat}{2}
&hhh: &\quad & v(\lambda_1-\lambda_2), \\
&hAA: &\quad & v(\lambda_1-\lambda_2), \\
&hHH: &\quad & v(\lambda_1-\lambda_2), \\
&hH^+H^-: &\quad &2v(\lambda_1+\lambda_2), \\
&hS^+S^-: &\quad &v\lambda_5, \\
&hS_1S_1: &\quad &\frac{v}{2}[\lambda_5+\lambda_6+2\cos\sigma\lambda_7], \\
&hS_2S_2: &\quad &\frac{v}{2}[\lambda_5+\lambda_6-2\cos\sigma\lambda_7], \\
&AS_1S_1: &\quad &v\sin\sigma \lambda_7, \\
&AS_2S_2: &\quad &-v\sin\sigma \lambda_7, \\
&HS_1S_2: &\quad &-2v\sin\sigma\lambda_7, \\
&H^+S^-S_1: &\quad &-iv\sin\sigma\lambda_7, \\
&H^+S^-S_2: &\quad &-v\sin\sigma\lambda_7.
\end{alignat}
\end{subequations}
We have here left out couplings involving the would-be Goldstone bosons.
Couplings involving $H^-S^+$ are obtained from those involving $H^+S^-$ by complex conjugation.
The $AAA$, $HHH$, $hhA$, $hhH$, $AAH$, $H^+H^-A$, $H^+H^-H$, $S^+S^-A$ and $S^+S^-H$ couplings all vanish.
The dependence on the phase $\sigma$ only appears in couplings involving $\lambda_7$ and two fields from the $S_3$ singlet sector.

%%%%%%%%%%%%%%%%%%%%%%%%%%%%%%%%%%%%%%%%%%%%%
\subsection{The phase \boldmath$\sigma$}
\label{Eq:subsect-sigma}
%%%%%%%%%%%%%%%%%%%%%%%%%%%%%%%%%%%%%%%%%%%%%
The phase $\sigma$ governs the mass splitting between the two neutral fields coming from the $S_3$ singlet Higgs doublet and it also appears in some of the trilinear couplings. This phase is always associated with $\lambda_7$
since in the C-III-c case this is the only coupling in the potential that is sensitive to a phase. Since the minimisation conditions do not constrain this phase, and this phase was also not included in the Lagrangian, it looks as if there are 
physical quantities that depend on parameters that are not physical. This apparent paradox can be solved by noticing that $\sigma$ can be promoted to a parameter of the potential by just rephasing the fields $h_1$ and $h_2$ in such a way that their vevs become real. What is special in this case is that there is a parameter that can either appear in the scalar potential or in the specification of the vacuum, depending on the choice of scalar basis. This parameter has physical implications and in particular contributes to the mass splitting of some scalars.
This corresponds to the transformation given by Eqs.~(\ref{def}).
The C-III-c vacuum differs from all other complex $S_3$ vacua, since in those cases the phases appear in the minimisation conditions and therefore cannot be considered as free parameters \cite{Emmanuel-Costa:2016vej}.

As mentioned above, in a basis where $\hat w_1=\hat w_2$, it is convenient to define $\sigma \equiv \sigma_1-\sigma_2$. We recall that this phase $\sigma$ is not determined by the potential. However, it parametrises correlations among certain physical couplings, and among those couplings and the mass splitting in the neutral $S_3$ singlet sector. The mass splitting and couplings given above, by Eqs.~(\ref{Eq:mass-splitting-singlet}) and (\ref{Eq:couplings-trilin}), assume a basis where $\hat w_1=\hat w_2=v/\sqrt2$.

Clearly,
\begin{equation}
\Delta^2\equiv m^2_{S_2}-m^2_{S_1}
\end{equation}
is a physical quantity, expressed as $2\cos\sigma\lambda_7 v^2$ in the adopted basis.
Next, if we denote the $AS_1S_1$ coupling ($v\sin\sigma\lambda_7$) by the abbreviation $\hat g$, then we can identify the modulus of $\lambda_7$ as a physical quantity:\footnote{It can be readily seen that this result does not depend on the basis chosen for the definitions of $H$ and $A$.}
\begin{equation}
|\lambda_7|=\sqrt{\left(\frac{\Delta^2}{2v^2}\right)^2+\left(\frac{\hat g}{v}\right)^2},
\end{equation}
whereas $\sigma$ (in the chosen basis) parametrises the ratio of the two physical quantities $\Delta^2$ and $\hat g$ by
\begin{equation}
\tan\sigma=\frac{2\hat gv}{\Delta^2}.
\end{equation}

%%%%%%%%%%%%%%%%%%%%%%%%%%%%%%%%%%%%%%%%%%%%%
\section{The softly broken potential} \label{sect:soft}
\setcounter{equation}{0}
%%%%%%%%%%%%%%%%%%%%%%%%%%%%%%%%%%%%%%%%%%%%%
We now replace the potential (\ref{Eq:pot-24}) by
\begin{equation}
V = V_2 + V_2^\prime + V_4,
\end{equation}
allowing for terms $V_2^\prime$ that softly break the $S_3$ symmetry.
The most general form of these terms is\footnote{Like for the unbroken case, we restrict ourselves to terms with real coefficients. The analysis of the $S_3$-symmetric model with complex parameters and with the addition of complex soft-breaking terms would give an important contribution to the understanding of such models.}
\begin{align}\label{VSoftGenericBasis}
V_2^\prime &=\mu_2^2 \left( h_1^\dagger h_1 - h_2^\dagger h_2 \right) 
+ \frac{1}{2} \nu_{12}^2 \left( h_1^\dagger h_2 + \mathrm{h.c.} \right)
+ \frac{1}{2} \nu_{01}^2 \left( h_S^\dagger h_1 + \mathrm{h.c.}\right)
+ \frac{1}{2} \nu_{02}^2 \left( h_S^\dagger h_2 + \mathrm{h.c.} \right).
\end{align}

The vacua studied in Ref.~\cite{Emmanuel-Costa:2016vej} will then be modified.
In the following we shall briefly discuss some general properties of the different vacua that result from the inclusion of soft $S_3$-breaking terms, employing a more generic terminology to label them. The labelling will specify how many and which vevs vanish, and our focus will be on massless states and mass degeneracies. Our approach is to fix the zero vevs in all possible positions and derive the resulting constraints.

%%%%%%%%%%%%%%%%%%%%%%%%%%%%%%%%%%%%%%%%%%%%%
\subsection{Real vacua}
%%%%%%%%%%%%%%%%%%%%%%%%%%%%%%%%%%%%%%%%%%%%%

We summarise in table~\ref{table:real_vevs} the different real vacua for the softly broken $S_3$-symmetric potential. In the following, we list some further properties, commenting also on the degeneracies that arise in the limit of no soft $S_3$-breaking terms. This classification is based on considering all vacua with two, or one or with no vanishing vevs, where the labels are self-explanatory. In table~\ref{Table:real} we did not include the case corresponding to $R_{x0y}$ because $w_2=0$ only appeared in the consistency conditions together with $w_S=0$. However, this could obviously be a limit of the general case R-III.

%%%%%%%%%%%%%%%%%%%%%%%%%%%%%%%%%%%%%%%%%%%%%
\begin{table}[htb]
\caption{Real vacua compatible with the most general soft $S_3$-breaking terms, Eq.~(\ref{VSoftGenericBasis}), together with the minimisation conditions.\label{table:real_vevs}}
\begin{center}
\begin{tabular}{|c|c|c|c|c|}
\hline
\hline
Label & $w_1,w_2,w_S$ & Constraints \\
\hline
\hline
$R_{00x}$ & $(0,0,w_S)$ & $\mu _0^2= -\lambda _8 w_S^2$, \\
&& $\nu_{01}^2=\nu_{02}^2=0$ \\
\hline
$R_{0x0}$ & $(0, w, 0)$ &$\mu _1^2=\mu_2^2 - \left(\lambda _1+\lambda _3\right) w^2$,\\
&&$\nu_{12}^2=0$, \quad $\nu_{02}^2=w^2\lambda_4$  \\
\hline
$R_{x00}$ & $(w,0,0)$ &  $\mu _1^2=-\mu_2^2- \left(\lambda _1+\lambda _3\right) w^2$,\\
&& $\nu_{12}^2=\nu_{01}^2=0$\\
\hline
\hline
$R_{0xy}$ & $(0,w,w_S)$ & $\mu _0^2=-\frac{1}{2}\nu_{02}^2\frac{ w}{w_S}+ \frac{1}{2}\lambda _4\frac{ w^3}{w_S}
-\frac{1}{2} \lambda_a w^2-\lambda _8 w_S^2$, \\
&& $\mu _1^2=\mu_2^2-\frac{1}{2}\nu_{02}^2\frac{w_S}{w} -\left( \lambda _1+ \lambda _3\right) w^2+\frac{3}{2} \lambda _4 w w_S-\frac{1}{2} \lambda_a w_S^2$,\\
&& $\nu_{12}^2=-\nu_{01}^2\frac{w_S}{w}$ \\
\hline
$R_{x0y}$ & $(w,0,w_S)$ & $\mu _0^2=-\frac{1}{2}\nu_{01}^2\frac{w}{w_S} -\frac{1}{2} \lambda_a w^2-\lambda _8 w_S^2$,\\
&& $\mu _1^2=-\mu_2^2-\frac{1}{2}\nu_{01}^2\frac{w_S}{w} - \left( \lambda _1+ \lambda _3\right) w^2-\frac{1}{2} \lambda_a w_S^2$, \\
&& $\nu_{12}^2=-\nu_{02}^2\frac{w_S}{w}-3\lambda_4ww_S$ \\
\hline
$R_{xy0}$ & $(w_1,w_2,0)$ & $\mu _1^2=-\nu_{12}^2\frac{w_1^2+w_2^2}{4w_1w_2} -\left(\lambda _1+\lambda _3\right)(w_1^2+w_2^2)$,\\
&& $\mu_2^2=\nu_{12}^2\frac{w_1^2-w_2^2}{4w_1w_2}$,\\
&& $\nu_{01}^2=\left[-\nu_{02}^2+(w_2^2-3w_1^2)\lambda_4\right]\frac{w_2}{w_1}$ \\
\hline
\hline
$R_{xyz}$ & $(w_1,w_2,w_S)$ & $\mu _0^2=-\frac{1}{2}\nu_{01}^2\frac{w_1}{w_S}-\frac{1}{2}\nu_{02}^2\frac{w_2}{w_S}-\frac{1}{2}\lambda_4\frac{w_2(3w_1^2-w_2^2)}{w_S} -\frac{1}{2} \lambda_a( w_1^2+ w_2^2)-\lambda _8 w_S^2$,\\
&& $\mu _1^2=-\nu_{12}^2\frac{w_1^2+w_2^2}{4w_1w_2} 
-\frac{w_S}{4w_1}\nu_{01}^2-\frac{w_S}{4w_2}\nu_{02}^2
-\frac{3}{4}\lambda_4\frac{(w_1^2+w_2^2)w_S}{w_2}$ \\
&& $- \left( \lambda _1+ \lambda _3\right)( w_1^2+ w_2^2)-\frac{1}{2} \lambda_a w_S^2$, \\
&& $\mu_2^2=\nu_{12}^2\frac{w_1^2-w_2^2}{4w_1w_2}-\frac{w_S}{4w_1}\nu_{01}^2+\frac{w_S}{4w_2}\nu_{02}^2-\frac{3}{4}\lambda_4\frac{w_S(3w_2^2-w_1^2)}{w_2}$ \\
\hline
\hline
\end{tabular}
\end{center}
\end{table}
%%%%%%%%%%%%%%%%%%%%%%%%%%%%%%%%%%%%%%%%%%%%%

Below, we briefly comment on some of the properties of the different categories of real vacua, allowing for soft breaking terms.

\subsubsection{\boldmath$(0,0,w_S)$}
Soft breaking terms $\nu_{01}^2$ and $\nu_{02}^2$ do not survive minimisation. If no soft breaking terms are present there is mass degeneracy among the charged scalars, as well as two pairs of mass degenerate neutral scalars. If either $\nu_{12}^2$ or $\mu_2^2$ is present there is no mass degeneracy.

\subsubsection{\boldmath$(0,w,0)$}
The soft breaking term $\nu_{12}^2$ does not survive minimisation. All masses are non-degenerate with or without soft breaking terms. If no soft breaking terms are present (R-II-2), this vacuum requires $\lambda_4=0$ and it has one massless state. If $\nu_{02}^2$ is the only soft breaking term, it requires $\lambda_4\neq0$, and we still have one neutral massless state. This massless state results from the condition relating $\nu_{02}^2$ and $\lambda_4$ in table~\ref{table:real_vevs}. If either $\mu_2^2$ or $\nu_{01}^2$ is present, then all neutral scalars become massive.

\subsubsection{\boldmath$(w,0,0)$}
Soft breaking terms $\nu_{12}^2$ and $\nu_{01}^2$ do not survive minimisation. There are no massless scalars and no mass degeneracy with or without the soft breaking terms.

\subsubsection{\boldmath$(0,w,w_S)$}
All soft breaking terms survive minimisation. There are no massless
scalars and no mass degeneracy with or without the soft breaking
terms. Note that the soft term $\nu_{12}^2$ and $\nu_{01}^2$ are proportional, i.e., they have to coexist.

\subsubsection{\boldmath$(w,0,w_S)$}
All soft breaking terms survive minimisation. There is no mass degeneracy with or without soft breaking terms. If no soft breaking term is present (requiring for consistency $\lambda_4=0$), one of the neutral masses vanishes. If there is at least one soft breaking term present, then all neutral scalars become massive.

\subsubsection{\boldmath$(w_1,w_2,0)$}
All soft breaking terms survive minimisation. There is no mass degeneracy with or without soft breaking terms. If no soft breaking term is present (requiring for consistency $\lambda_4=0$), one of the neutral masses vanishes. If there is at least one soft breaking term present, then all neutral scalars become massive.
Note that the $\mu_2^2$ term cannot be the only soft term.

\subsubsection{\boldmath$(w_1,w_2,w_S)$}
All soft breaking terms survive minimisation. There is no mass degeneracy with or without soft breaking terms. If no soft breaking term is present (requiring for consistency $\lambda_4=0$), one of the neutral masses vanishes. If there is at least one soft breaking term present, then all neutral scalars become massive.

%%%%%%%%%%%%%%%%%%%%%%%%%%%%%%%%%%%%%%%%%%%%%
\subsection{Complex vacua}
%%%%%%%%%%%%%%%%%%%%%%%%%%%%%%%%%%%%%%%%%%%%%

Below, we briefly comment on some of the properties of the different categories of complex vacua, allowing for soft breaking terms.

%%%%%%%%%%%%%%%%%%%%%%%%%%%%%%%%%%%%%%%%%%%%%
\begin{table}
\caption{Complex vacua compatible with the most general soft $S_3$-breaking terms, Eq.~(\ref{VSoftGenericBasis}), together with the minimisation conditions. The following abbreviations are introduced: $S_\pm = \sin^2\sigma_1\hat w_1^2 \pm \sin^2\sigma_2\hat w_2^2$.\label{table:complex_vevs}}
\begin{center}
\resizebox{\textwidth}{!}{
\begin{tabular}{|c|c|}
\hline
\hline
 $w_1,w_2,w_S$ & Constraints \\
\hline
\hline
\begin{tabular}[c]{@{}c@{}} $(0,\hat w_2e^{i\sigma_2},\hat w_S)$\\ \vspace{-10pt}  \\ $C_{0xy}$\end{tabular} & 
\begin{tabular}[c]{@{}c@{}} $\mu _0^2=- \frac{1}{2} \lambda_b \hat{w}_2^2-\lambda _8 \hat{w}_S^2$,\\
$\mu _1^2=\mu_2^2-\left(\lambda _1+\lambda _3\right) \hat{w}_2^2+\cos\sigma_2\hat w_2\hat w_S\lambda_4- \frac{1}{2} \lambda _b\hat{w}_S^2$, \\
$\nu_{02}^2=\hat w_2(\hat w_2\lambda_4-4\cos\sigma_2\hat w_S\lambda_7)$, \\
$\nu_{12}^2=\nu_{01}^2=0$ \end{tabular}\\
\hline
\begin{tabular}[c]{@{}c@{}} $(\hat w_1e^{i\sigma_1},0,\hat w_S)$ \\ $C_{x0y}$  \end{tabular} & 
\begin{tabular}[c]{@{}c@{}} $\mu _0^2=-\frac{1}{2} \lambda_b \hat{w}_1^2-\lambda _8 \hat{w}_S^2$,\\
$\mu _1^2=-\mu_2^2-\left( \lambda _1+ \lambda _3\right) \hat{w}_1^2-\frac{1}{2} \lambda_b \hat{w}_S^2$, \\
$\nu_{12}^2=-2\cos\sigma_1\hat w_1\hat w_S\lambda_4,$ \\ 
$\nu_{01}^2=-4\cos\sigma_1\hat w_1\hat w_S\lambda_7$ \\
$\nu_{02}^2=-\lambda_4\hat w_1^2$ \end{tabular}\\
\hline
\begin{tabular}[c]{@{}c@{}}  $(\hat w_1 e^{i\sigma_1},\hat w_2e^{i\sigma_2},0)$ \\ $C_{xy0}$\end{tabular} & 
\begin{tabular}[c]{@{}c@{}} $\mu_1^2=-(\lambda_1-\lambda_2)(\hat{w}_1^2+\hat{w}_2^2)$,\\
$\mu_2^2=-(\hat w_1^2-\hat w_2^2)(\lambda_2+\lambda_3)$, \\
$\nu_{12}^2=-4\hat w_1\hat w_2\cos(\sigma_2-\sigma_1)(\lambda_2+\lambda_3)$,\\
$\nu_{01}^2=-2\hat w_1\hat w_2\cos(\sigma_2-\sigma_1)\lambda_4$,\\
$\nu_{02}^2=-(\hat w_1^2-\hat w_2^2)\lambda_4$ \end{tabular}\\
\hline
\begin{tabular}[c]{@{}c@{}} $(\hat w_1 e^{i\sigma_1},\hat w_2e^{i\sigma_2},\hat w_S)$ \\ $C_{xyz}$\\
\\
provided\\
$\sin(\sigma_2-\sigma_1)\neq0.$
\end{tabular}&
\begin{tabular}[c]{@{}c@{}} 
$\mu_1^2=\{2\hat w_S^2S_+\mu_0^2-4\hat w_1^2\hat w_2^2(\hat w_1^2+\hat w_2^2)\sin^2(\sigma_2-\sigma_1)(\lambda_1-\lambda_2)$ \\ 
$-\hat w_S^2[2\hat w_1^2\hat w_2^2\sin^2(\sigma_2-\sigma_1)
-(\hat w_1^2+\hat w_2^2)S_+]\lambda_b +2\hat w_S^4S_+\lambda_8\}/$ \\ \vspace{6pt}
$[4\hat w_1^2\hat w_2^2\sin^2(\sigma_2-\sigma_1)]$ \\ 
$\mu_2^2=-\{2\hat w_S^2S_-\mu_0^2+4\hat w_1^2\hat w_2^2(\hat w_1^2
-\hat w_2^2)\sin^2(\sigma_2-\sigma_1)(\lambda_2+\lambda_3) $ \\ 
$+4\cos\sigma_2\hat w_1^2\hat w_2^3\hat w_S\sin^2(\sigma_2-\sigma_1)\lambda_4$\\ \vspace{6pt}
$+\hat w_S^2S_-[(\hat w_1^2+\hat w_2^2)\lambda_b +2\hat w_S^2\lambda_8]\}/
[4\hat w_1^2\hat w_2^2\sin^2(\sigma_2-\sigma_1)]$ \\ 
$\nu_{12}^2=-\{2\sin\sigma_1\sin\sigma_2\hat w_S^2\mu_0^2 
+4\hat w_1^2\hat w_2^2\cos (\sigma_2-\sigma_1) \sin^2 (\sigma_2-\sigma_1)(\lambda_2+\lambda_3)$\\
$+2\cos\sigma_1\hat w_1^2\hat w_2\hat w_S\sin^2(\sigma_2-\sigma_1)\lambda_4
+\sin\sigma_1\sin\sigma_2(\hat w_1^2+\hat w_2^2)\hat w_S^2\lambda_b$ \\  \vspace{6pt}
$+2\sin\sigma_1\sin\sigma_2\hat w_S^4\lambda_8
\}/[\hat w_1\hat w_2\sin^2(\sigma_2-\sigma_1)]$ \\
$\nu_{01}^2=-\{2\sin\sigma_2\hat w_S\mu_0^2
+\hat w_1^2\hat w_2\sin[2(\sigma_2-\sigma_1)]\lambda_4
+\sin\sigma_2\hat w_S(\hat w_1^2+\hat w_2^2)\lambda_b$ \\ \vspace{6pt}
$+4\cos\sigma_1\hat w_1^2\hat w_S\sin(\sigma_2-\sigma_1)\lambda_7
+2\sin\sigma_2\hat w_S^3\lambda_8
\}/[\hat w_1\sin(\sigma_2-\sigma_1)]$, \\
 $\nu_{02}^2=\{2\sin\sigma_1 \hat w_S\mu_0^2
+\hat w_2(\hat w_2^2-\hat w_1^2)\sin(\sigma_2-\sigma_1)\lambda_4$
$+\sin\sigma_1\hat w_S(\hat w_1^2+\hat w_2^2)\lambda_b $ \\
$-4\cos\sigma_2\hat w_2^2\hat w_S\sin(\sigma_2-\sigma_1)\lambda_7
+2\sin\sigma_1\hat w_S^3\lambda_8
\}/[\hat w_2\sin(\sigma_2-\sigma_1)]$ \end{tabular}\\
\hline
\begin{tabular}[c]{@{}c@{}} $(\hat w_1 e^{i\sigma},\pm\hat w_2e^{i\sigma},\hat w_S)$ \\ $C_{xyz}$\\
	\\
	provided\\
	$\sin\sigma\neq0.$
\end{tabular}&
\begin{tabular}[c]{@{}c@{}} 
	$\mu_0^2  =-\frac{1}{2}  \left(\hat{w}_1^2+\hat{w}_2^2\right)\lambda_b -\lambda_8 \hat{w}_S^2$,\\
	$\mu_1^2  = \frac{-1}{4 \hat{w}_1 \hat{w}_2} \{ 4\hat w_1\hat w_2(\hat w_1^2+\hat w_2^2)(\lambda_1+\lambda_3)
	+2 \hat{w}_1 \hat{w}_S \left[\left(\hat{w}_1^2+\hat{w}_2^2\right) \cos\sigma\lambda_4
	+\hat{w}_2 \hat{w}_S\lambda_b\right]$\\
	$\pm(\hat w_1^2+\hat w_2^2)\nu_{12}^2 \}$,\\
	$\mu_2^2  = \frac{1}{4 \hat{w}_1 \hat{w}_2} \{ 2 \lambda_4 \hat{w}_1 \hat{w}_S \left(\hat{w}_1^2-3 \hat{w}_2^2\right)  \cos\sigma\pm\left(\hat{w}_1^2-\hat{w}_2^2\right)\nu_{12}^2  \}$,\\
	$\nu_{01}^2  = \mp2 \hat{w}_1 (\lambda_4 \hat{w}_2+2 \lambda_7 \hat{w}_S \cos\sigma)$,\\
	$\nu_{02}^2  = \lambda_4 \left(\hat{w}_2^2-\hat{w}_1^2\right)-4 \lambda_7 \hat{w}_2 \hat{w}_S \cos\sigma$.
\end{tabular}\\
\hline
\hline
\end{tabular}}
\end{center}
\end{table}
%}
%%%%%%%%%%%%%%%%%%%%%%%%%%%%%%%%%%%%%%%%%%%%%

%\clearpage

\subsubsection{\boldmath$(0,\hat w_2e^{i\sigma_2},\hat w_S)$}
Soft breaking terms $\nu_{12}^2$ and $\nu_{01}^2$ do not survive minimisation. There are no massless scalars and no mass degeneracy with or without the soft breaking terms.

\subsubsection{\boldmath$(\hat w_1e^{i\sigma_1},0,\hat w_S)$}
All soft breaking terms survive minimisation. 

Let us first assume $\sigma_1\neq\pm\frac{\pi}{2}$ (C-IV-a).
If there are no soft breaking terms (requiring $\lambda_4=\lambda_7=0$) we have two massless neutral scalars. 
With $\lambda_4$ and $\lambda_7$ both equal to zero, the scalar potential acquires an $\text{O(2)}\otimes\text{U(1)}_{h_S}$ symmetry, where $\text{U(1)}_{h_S}$ refers to the independent rephasing of $h_S$. Vacua that break these two continuous symmetries lead to two massless neutral scalars.

It is impossible to have $\nu_{12}^2$ or $\nu_{02}^2$ as the only soft breaking term.
If the $\mu_2^2$  term is the only soft breaking term, we have one massless neutral scalar.
If the $\nu_{01}^2$ term is present, there are no massless neutral scalars.

Possible situations where we have only two soft breaking terms are when we have $\nu_{12}^2$ and $\nu_{02}^2$ (this situation requires $\lambda_4\neq0$) or when we have $\mu_2^2$ and $\nu_{01}^2$ (this situation requires $\lambda_7\neq0$). In both these situations there are no massless neutral scalars. 

Possible situations where we have three soft breaking terms are when we have $\nu_{12}^2$, $\mu_2^2$ and $\nu_{02}^2$ or when we have $\nu_{12}^2$, $\nu_{01}^2$ and $\nu_{02}^2$. In either of these two situations there are no massless neutral scalars. If all four soft breaking terms are present there are no massless neutral scalars.

Next, let us assume $\sigma_1=\pm\frac{\pi}{2}$ (C-III-b).
In this case we immediately get $\nu_{12}^2=\nu_{01}^2=0$, so the only possible soft breaking terms are $\mu_2^2$ and $\nu_{02}^2$. If there are no soft breaking terms (this requires $\lambda_4=0$) we have one massless scalar. If the $\mu_2^2$ or $\nu_{02}^2$ term is present, there are no massless scalars.

\subsubsection{\boldmath$(\hat w_1e^{i\sigma_1},\hat w_2e^{i\sigma_2},0)$}
All soft breaking terms survive minimisation. 

Let us first assume $\sigma_1-\sigma_2\neq\pm\frac{\pi}{2}$ (C-III-c).
If there are no soft breaking terms (this requires $\lambda_2+\lambda_3=0$ and $\lambda_4=0$) we have two massless neutral scalars. Not all combinations of soft breaking terms are allowed, but if at least one soft breaking term is present we have no massless neutral scalars.

Next, let us assume $\sigma_1-\sigma_2=\pm\frac{\pi}{2}$. In this case we immediately get $\nu_{12}^2=\nu_{01}^2=0$, so the only possible soft breaking terms are $\mu_2^2$ and $\nu_{02}^2$. If $\hat w_2 \neq \pm \hat w_1$ and no soft breaking terms are present, we have two massless scalars. If either of the soft breaking terms are present, we have no massless neutral scalars.

Finally, if $\sigma_1-\sigma_2=\pm\frac{\pi}{2}$ and $\hat w_2 = \pm \hat w_1$ (C-I-a), there are no soft breaking terms and also no massless neutral scalars.

\subsubsection{\boldmath$(\hat w_1e^{i\sigma_1},\hat w_2e^{i\sigma_2},\hat w_S)$}
All four soft breaking terms survive minimisation.
The case $\sin(\sigma_2-\sigma_1)=0$ requires special attention, and is listed separately in table~\ref{table:complex_vevs}.
Finally, if $\sigma_1-\sigma_2\neq0$, then all soft terms are constrained by the parameters of the unbroken potential, together with the vevs (moduli $\hat w_1$, $\hat w_2$, $\hat w_S$, and the phases $\sigma_1$ and $\sigma_2$).

%%%%%%%%%%%%%%%%%%%%%%%%%%%%%%%%%%%%%%%%%%%%%
\section{The C-III-c model with soft \boldmath$S_3$-breaking} \label{sect:C-III-c-soft}
\setcounter{equation}{0}
%%%%%%%%%%%%%%%%%%%%%%%%%%%%%%%%%%%%%%%%%%%%%

The C-III-c vacuum was characterised \cite{Emmanuel-Costa:2016vej} as a vacuum with $\hat w_S=0$ and with the other two vevs arbitrary complex, $(\hat w_1e^{i\sigma_1},\hat w_2e^{i\sigma_2},0)$. It is worth stressing that the C-I-a vacuum ($\hat w_1,\pm i\hat w_1,0$) is not a particular case of C-III-c. This is clearly seen by comparing the constraints arising from the stationary-point equations. Whereas there are three constraints attached to C-III-c, see Eq.~(\ref{Eq:C-III-c-conditions}), there is only one constraint attached to C-I-a, namely
$\mu _1^2=-\left(\lambda _1-\lambda _2\right) v^2$.
The C-III-c vacuum requires $\lambda_4=0$, $\lambda_2+\lambda_3=0$, and $\mu_1^2=-(\lambda_1-\lambda_2) v^2$ and can in fact be simplified to $(\hat we^{i\sigma},\hat w,0)$ due to the O(2) symmetry resulting from $\lambda_4=0$.

The C-I-a case was studied by Derman and Tsao \cite{Derman:1979nf} and by Branco, G\'erard and Grimus \cite{Branco:1983tn}. It has the property of being geometrical in the sense that it is complex with the phases fixed by the symmetry rather than by the parameters of the potential. In the defining representation this vacuum appears as ($x,xe^{2\pi i/3},x^{-2\pi i/3}$). It was shown \cite{Branco:1983tn} that this vacuum does not violate CP in spite of being complex.

One of the constraints of C-III-c is $\lambda_4=0$. Whenever $\lambda_4=0$ the potential acquires an additional O(2) symmetry which is continuous. This symmetry is broken by the vevs and therefore there will be massless scalars. There will in fact be two massless scalars, as discussed above. One way of avoiding massless scalars is to include soft breaking terms. Soft breaking terms combining $h_S$ with one of the $h_i$ are not consistent with $\lambda_4=0$.

It was shown \cite{Emmanuel-Costa:2016vej,Ogreid:2017alh} that the presence of the O(2) symmetry allows one to transform the C-III-c vacuum into ($\hat we^{i\sigma/2},\hat w^{-i\sigma/2},0$) and therefore it can be readily shown that it also preserves CP.
Actually, the introduction of soft breaking terms can only introduce CP violation in the mass part of the potential, not in the interactions. Thus, if the mass-squared matrices split into a CP-even and a CP-odd part, then CP is conserved.

The C-III-c vacuum with soft breaking terms is denoted as $C_{xy0}$ in table~\ref{table:complex_vevs}. The minimisation constraints are
\begin{subequations} \label{Eq:minimisation-soft}
\begin{align}
\mu_1^2&=-(\hat{w}_1^2+\hat{w}_2^2)(\lambda_1-\lambda_2), \\
\mu_2^2&=-(\hat w_1^2-\hat w_2^2)(\lambda_2+\lambda_3), \\
\nu_{12}^2&=-4\hat w_1\hat w_2\cos(\sigma_2-\sigma_1)(\lambda_2+\lambda_3), \label{Eq:nu_12}\\
\nu_{01}^2&=-2\hat w_1\hat w_2\cos(\sigma_2-\sigma_1)\lambda_4, \label{Eq:nu_01}\\
\nu_{02}^2&=-(\hat w_1^2-\hat w_2^2)\lambda_4. \label{Eq:nu_02}
\end{align}
\end{subequations}
It is clear from these expressions that for $\mu_{2}^2$ and $\nu_{12}^2$ can only be different from zero if $\lambda_2+\lambda_3\neq0$. Likewise, $\nu_{01}^2$ and $\nu_{02}^2$ can only be present for $\lambda_4\neq0$.

We shall here present three avenues to the introduction of modifications to C-III-c:
\begin{itemize}
\item
Models with vevs $(\hat w_1 e^{i\sigma_1},\hat w_2e^{i\sigma_2},0)$, $\lambda_4=0$ and $\lambda_2+\lambda_3\neq0$,
\item
Models with vevs $(\hat w_1 e^{i\sigma_1},\hat w_2e^{i\sigma_2},0)$, $\lambda_4\neq0$ and $\lambda_2+\lambda_3=0$,
\item
Models with vevs $(\hat w_1 e^{i\sigma_1},\hat w_2e^{i\sigma_2},0)$, $\lambda_4\neq0$ and $\lambda_2+\lambda_3\neq0$.
\end{itemize}

%%%%%%%%%%%%%%%%%%%%%%%%%%%%%%%%%%%%%%%%%%%%%
\begin{table}[htb]
\caption{Summary of softly-broken C-III-c-like vacua. Here, ``SBT'' stands for ``Soft-breaking terms''.
When the two moduli are equal, we denote it $\hat w$.
In the last column we listed the symmetry responsible for no spontaneous CP violation.}
\label{Table:soft-summary}
\begin{center}
%\resizebox{0.97\textwidth}{!}{
\begin{tabular}{|c|c|c|c|c|c|c|}
\hline\hline
Case &Constraints & Allowed SBT & Vacuum & CP  \\
\hline
\hline
1 & $\lambda_4=0$, $\lambda_2+\lambda_3=0$ & none & $(\hat w_1e^{i\sigma_1},\hat w_2e^{i\sigma_2},0)$ & conserving \\
& C-III-c & & $\equiv(\hat w e^{i\sigma/2},\hat we^{-i\sigma/2},0)$ & O(2) \\
\hline
2 & $\lambda_4=0$, $\lambda_2+\lambda_3\neq0$ & $\mu_2^2$ & $(\pm i\hat w_1,\hat w_2,0)$ & conserving \\
& $\cos(\sigma_2-\sigma_1)=0$, $\hat w_1\neq\hat w_2$ &  &  & $h_1\to-h_1$ \\
\hline
3 & $\lambda_4=0$, $\lambda_2+\lambda_3\neq0$ & $\nu_{12}^2$ & $(\hat we^{i\sigma_1},\hat we^{i\sigma_2},0)$ & conserving \\
& $\cos(\sigma_2-\sigma_1)\neq0$, $\hat w_1=\hat w_2$ &  & $\equiv(\hat w e^{i\sigma/2},\hat we^{-i\sigma/2},0)$ & $h_1\leftrightarrow h_2$ \\
\hline
4 & $\lambda_4=0$, $\lambda_2+\lambda_3\neq0$ & $\mu_2^2$, $\nu_{12}^2$ & $(\hat w_1e^{i\sigma_1},\hat w_2e^{i\sigma_2},0)$ & conserving \\
& no other conditions &  &  &  \\
\hline
5 & $\lambda_4\neq0$, $\lambda_2+\lambda_3=0$ & none & $(\pm i\hat w,\hat w,0)$ & conserving \\
& $\cos(\sigma_2-\sigma_1)=0$, $\hat w_1=\hat w_2$ &  &  & $h_1\to-h_1$ \\
& C-I-a &  &  &  \\
\hline
6 & $\lambda_4\neq0$, $\lambda_2+\lambda_3=0$ & $\nu_{02}^2$ & $(\pm i\hat w_1,\hat w_2,0)$ & conserving \\
& $\cos(\sigma_2-\sigma_1)=0$, $\hat w_1\neq\hat w_2$ &  &  & $h_1\to-h_1$ \\
\hline
7 & $\lambda_4\neq0$, $\lambda_2+\lambda_3=0$ & $\nu_{01}^2$ & $(\hat we^{i\sigma},\hat w,0)$ & violating \\
& $\cos(\sigma_2-\sigma_1)\neq0$, $\hat w_1=\hat w_2$ &  &  &  \\
\hline
8 & $\lambda_4\neq0$, $\lambda_2+\lambda_3=0$ & $\nu_{01}^2$, $\nu_{02}^2$ & $(\hat w_1e^{i\sigma_1},\hat w_2e^{i\sigma_2},0)$ & violating \\
& no other conditions &  &  &  \\
\hline
9 & $\lambda_4\neq0$, $\lambda_2+\lambda_3\neq0$ & $\mu_2^2$, $\nu_{02}^2$ & $(\pm i\hat w_1,\hat w_2,0)$ & conserving \\
& $\cos(\sigma_2-\sigma_1)=0$, $\hat w_1\neq\hat w_2$ &  &  & $h_1\to-h_1$  \\
\hline
10 & $\lambda_4\neq0$, $\lambda_2+\lambda_3\neq0$ & $\nu_{12}^2$, $\nu_{01}^2$ & $(\hat we^{i\sigma},\hat w,0)$ & violating \\
& $\cos(\sigma_2-\sigma_1)\neq0$, $\hat w_1=\hat w_2$ &  &  &  \\
\hline
11 & $\lambda_4\neq0$, $\lambda_2+\lambda_3\neq0$ & all & $(\hat w_1e^{i\sigma_1},\hat w_2e^{i\sigma_2},0)$ & violating \\
& $\sigma_2-\sigma_1\neq0$, $\hat w_1\neq\hat w_2$ &  &  &  \\
\hline
\hline
\end{tabular}
%}
\end{center}
\end{table}
%%%%%%%%%%%%%%%%%%%%%%%%%%%%%%%%%%%%%%%%%%%%%
%%%%%%%%%%%%%%%%%%%%%%%%%%%%%%%%%%%%%%%%%%%%%
\subsection{Models with vevs \boldmath$(\hat w_1 e^{i\sigma_1},\hat w_2e^{i\sigma_2},0)$, $\lambda_4=0$ and $\lambda_2+\lambda_3\neq0$}
\label{sect:lam4=0}
%%%%%%%%%%%%%%%%%%%%%%%%%%%%%%%%%%%%%%%%%%%%%

It is clear from expressions (\ref{VSoftGenericBasis}), (\ref{Eq:nu_01}) and (\ref{Eq:nu_02}) that the condition $\lambda_4=0$ is not consistent with having soft breaking terms involving $h_S$ and either one of the $h_i$. Soft breaking terms involving $h_1$ and $h_2$ are only possible if we relax the condition $\lambda_2+\lambda_3=0$.

The introduction of soft breaking terms of $S_3$ also breaks the O(2) symmetry that resulted from having $\lambda_4$ equal to zero. As a result, in this case we cannot use this symmetry to write this set of vacua with equal moduli for the first two entries. 

In general, CP is spontaneously broken in this case, provided that $\cos(\sigma_2-\sigma_1)\neq0$. 
It can be readily seen that, if $\cos(\sigma_2-\sigma_1)=0$, then $\nu_{12}^2$ must be zero and only the soft breaking term proportional to $\mu_2^2$ survives. The vacuum will have the form $(\pm i\hat w_1,\hat w_2,0)$, the initial symmetry $h_1\to-h_1$ is not broken and CP is conserved by the vacuum, since the following condition \cite{Branco:1983tn}
\begin{equation} \label{Eq:BGG}
 U_{ij} \langle 0| \Phi_j |0\rangle^\ast = \langle 0| \Phi_i |0\rangle
\end{equation}
is satisfied for
\begin{equation} \label{Eq:first_vac_case}
U=
\begin{pmatrix}
-1 & 0 & 0\\
0 & 1 & 0\\
0 & 0 & 1
\end{pmatrix}.
\end{equation}

It can be checked that by going to the Higgs basis, according to the method proposed in Ref.~\cite{Ogreid:2017alh}, that  CP is conserved when the following four conditions are met: 
\begin{subequations} \label{Eq:first_case}
\begin{alignat}{2}
&\bullet &\quad &\sin[2(\sigma_2-\sigma_1)]=0, \label{Eq_lam23}\\\
&\bullet &\quad &\lambda_7(\sin2\sigma_1-\sin2\sigma_2)=0, \label{Eq_lam7-1}\\
&\bullet &\quad &\lambda_7(\hat w_1^2\sin2\sigma_1+\hat w_2^2\sin2\sigma_2)=0, \label{Eq_lam7-2}\\
&\bullet &\quad &\lambda_7(\hat w_1^2\sin2\sigma_2+\hat w_2^2\sin2\sigma_1)=0. \label{Eq_lam7-3}
\end{alignat}
\end{subequations}
Clearly, if $\sigma_1=\sigma_2=0$ there is no CP violation. Actually, these conditions are {\it sufficient}, not {\it necessary} for CP conservation. This distinction will be illustrated by Case~4 discussed in appendices~\ref{app:B} and \ref{app:C}.

If $\cos(\sigma_2-\sigma_1)\neq0$ and $\hat w_1=\hat w_2$ the term in $\mu_2^2$  is forced to be zero and only the soft breaking term in $\nu_{12}^2$ survives. The vacuum will have the form $(\hat we^{i\sigma_1},\hat we^{i\sigma_2},0)$ which can be rephased into $(\hat we^{i\sigma},\hat we^{-i\sigma},0)$. The potential will have symmetry for $h_1\leftrightarrow h_2$ and CP is conserved with the following choice of $U$ in Eq.~(\ref{Eq:BGG})
\begin{equation}
U=\begin{pmatrix}
 0 & 1 & 0\\
1 & 0 & 0\\
0 & 0 & 1
\end{pmatrix}.
\end{equation}

In Case 4 of table~\ref{Table:soft-summary} one might expect CP to be violated. However, this is not the case as will be shown in appendix~\ref{app:C}. In order to prove it one can go to the Higgs basis where only one of the fields acquires a non-zero real vev and use the freedom to rephase the fields with zero vev in order to make all the coefficients of the potential real \cite{Ogreid:2017alh}. This set of transformations changes the form of the potential but does not change the physics. In its final version both the potential and the vevs are real.

%%%%%%%%%%%%%%%%%%%%%%%%%%%%%%%%%%%%%%%%%%%%%
\subsection{Models with vevs \boldmath$(\hat w_1 e^{i\sigma_1},\hat w_2e^{i\sigma_2},0)$, $\lambda_4\neq0$ and $\lambda_2+\lambda_3=0$}
\label{sect:lam2+lam3=0}
%%%%%%%%%%%%%%%%%%%%%%%%%%%%%%%%%%%%%%%%%%%%%
These are also models denoted by $C_{xy0}$ in table~\ref{table:complex_vevs}. 
In the particular case of $\cos(\sigma_2-\sigma_1)=0$ and $\hat w_1=\hat w_2$, no soft breaking term survives and we fall into the case C-I-a and there is no spontaneous CP violation.

We now find that CP is violated unless Eqs.~(\ref{Eq_lam7-1})--(\ref{Eq_lam7-3}) are satisfied, together with:
\begin{subequations} \label{Eq:second_case}
\begin{alignat}{2}
&\bullet &\quad &\hat w_1^2\sin(2\sigma_1-\sigma_2)+(2\hat w_1^2-\hat w_2^2)\sin\sigma_2=0, \\
&\bullet &\quad &\hat w_2^2\sin(2\sigma_1-\sigma_2)+(2\hat w_2^2-\hat w_1^2)\sin\sigma_2=0, \\
&\bullet &\quad &\hat w_1^2\sin(2\sigma_1-\sigma_2)-3\hat w_2^2\sin\sigma_2=0, \\
&\bullet &\quad &\hat w_2^2\sin(2\sigma_1-\sigma_2)-3\hat w_1^2\sin\sigma_2=0,
\end{alignat}
\end{subequations}
which replace Eq.~(\ref{Eq_lam23}).
Clearly, if $\sigma_1=\sigma_2=0$ there is no CP violation. There are two special cases worth considering:
\begin{itemize}
\item
$\cos(\sigma_2-\sigma_1)=0$, and $\hat w_1\neq\hat w_2$,
\item
$\cos(\sigma_2-\sigma_1)\neq0$, and $\hat w_1=\hat w_2$.
\end{itemize}
In the first case, the vacuum can be written as $(\pm i\hat w_1,\hat w_2,0)$, where we chose $\sigma_2=0$ by the freedom to rephase. In this case, the term proportional to $\nu_{01}^2$ does not survive, and the potential is symmetric under $h_1\to-h_1$. CP is therefore conserved since Eq.~(\ref{Eq:first_vac_case}) is satisfied.

In the second case, only the term proportional to $\nu_{01}^2$ survives, and in general CP will be violated (unless both phases vanish). This phenomenon illustrates the fact that soft breaking terms can introduce spontaneous CP violation \cite{Branco:1985aq}. 

%%%%%%%%%%%%%%%%%%%%%%%%%%%%%%%%%%%%%%%%%%%%%
\subsection{Models with vevs \boldmath$(\hat w_1 e^{i\sigma_1},\hat w_2e^{i\sigma_2},0)$, $\lambda_4\neq0$ and $\lambda_2+\lambda_3\neq0$}
\label{sect:lam2+lam3neq0_and_lam4neq0}
%%%%%%%%%%%%%%%%%%%%%%%%%%%%%%%%%%%%%%%%%%%%%

In this general case, going to the Higgs basis, we see that the full set of conditions, including Eqs.~(\ref{Eq:first_case}) and (\ref{Eq:second_case}), must be satisfied. Obviously, for $\sigma_2-\sigma_1=0$, these equations can be verified after a suitable rephasing.
Otherwise, CP is violated.

Table~\ref{Table:soft-summary} summarises the results obtained in this section.
In Case 3, 7 and 10 the vacuum can be written in the same form, but Case 3 is CP conserving, while the other two are CP violating.

In appendix~\ref{app:B} we collect some information on the mass spectra of these different models.

%%%%%%%%%%%%%%%%%%%%%%%%%%%%%%%%%%%%%%%%%%%%%
\section{Summary}
\setcounter{equation}{0}
%%%%%%%%%%%%%%%%%%%%%%%%%%%%%%%%%%%%%%%%%%%%%
We have presented a detailed discussion of some 3HDM vacua obtained from an $S_3$-symmetric potential with soft symmetry breaking. The $S_3$-symmetric potential, for certain vev alignments, is plagued with massless states and degeneracies, whose origins have been identified. In fact, all possible vacua, except $(x,x,x)$, expressed in the defining representation, break the $S_3$
symmetry spontaneously. We have shown that some vacuum solutions require conditions on the parameters of the potential that lead to accidental continuous symmetries which in turn are broken by the vacuum. Allowing for soft $S_3$-breaking terms, all states become massive.

The case of a vanishing singlet vev, without the introduction of soft breaking terms, which we denoted as C-III-c is particularly interesting. It exhibits an unfamiliar feature: the minimum of the potential allows for a relative phase between the two non-zero vevs, whose value is not constrained by the minimisation conditions \cite{Emmanuel-Costa:2016vej}. For all other complex $S_3$ vacua the phases always appear in the minimisation conditions and therefore cannot be considered as free parameters \cite{Emmanuel-Costa:2016vej}. In the C-III-c case the phase appearing in the vevs is an additional free parameter which determines the mass splitting in the neutral $S_3$ singlet sector. This phase also shows up in certain couplings. Whenever this phase is chosen to be zero the two neutral scalars from the $S_3$ singlet sector are degenerate in mass. It is possible to remove this phase from the vevs by a rephasing giving rise to a CP conserving potential with $\lambda_7$ complex. In this basis all free parameters appear in the potential.

This work focuses on the C-III-c vacuum as well as on its versions with soft breaking terms. It is beyond the scope of this paper to discuss other possible vacua with soft breaking terms. However, models with $\hat w_S$ non-zero may also provide interesting possibilities from the phenomenological point of view.

Furthermore, we have seen that the form of the vacuum does not determine whether or not CP is violated spontaneously. One and the same form of the vacuum may conserve or violate CP, depending on which soft $S_3$-breaking terms (and corresponding constraints) are present.

The results presented in this paper should be useful for model building, providing guidelines for
various interesting scenarios. The $S_3$-symmetric potential with three Higgs doublets has been analysed by several authors in the past few years with many different 
aims \cite{Chakrabarty:2015kmt}, such as looking for realistic Yukawa couplings \cite{Das:2015sca, Gomez-Izquierdo:2018jrx, Chakrabarty:2019tsm, Das:2019yad}, looking for dark matter candidates
\cite{Machado:2012ed,Fortes:2014dca,Espinoza:2018itz,Mishra:2019keq}, looking for CP violation \cite{Gerard:1982mm,Barradas-Guevara:2015rea}, as well as many other studies.

%\clearpage

\section*{Acknowledgements}
It is a pleasure to thank Pedro Ferreira, Igor Ivanov and Dieter L\"ust for discussions on the massless states.
We also thank the referee for raising some points that allowed us to clarify some important details.
PO and MNR thank the CERN Theory Division, where part of this work was done, for hospitality.
PO~is supported by the Research Council of Norway.
The work of MNR was partially supported by Funda\c c\~ ao 
para  a  Ci\^ encia e a Tecnologia  (FCT, Portugal)  through  the  projects  
CFTP-FCT Unit  777(UID/FIS/00777/2013), (UID/FIS/00777/2019), CERN/FIS-PAR/0004/2017 and 
PTDC/FIS-PAR/29436/2017  which are  partially  funded  through
POCTI  (FEDER),  COMPETE,  QREN  and  EU. OMO, PO and MNR benefited from discussions that
took place at the University of Warsaw during visits supported by the HARMONIA project of the National Science Centre,
Poland, under contract UMO-2015/18/M/ST2/00518 (2016--2019).
MNR and PO also thank the University of Bergen  and
CFTP/IST/University of Lisbon, where collaboration visits took place. 
%\clearpage

\appendix
%%%%%%%%%%%%%%%%%%%%%%%%%%%%%%%%%%%%%%%%%%%%%
\section{Real and complex vacua of the \boldmath$S_3$-symmetric potential}
\setcounter{equation}{0}
\label{app:A}
%%%%%%%%%%%%%%%%%%%%%%%%%%%%%%%%%%%%%%%%%%%%%

For convenience, we include tables~\ref{Table:real} and \ref{Table:complex} that summarise some of the properties of the possible real and complex vacuum solutions \cite{Emmanuel-Costa:2016vej}.

%%%%%%%%%%%%%%%%%%%%%%%%%%%%%%%%%%%%%%%%%%%%%
\begin{table}[htb]
\caption{Possible real vacua (partly after Derman and Tsao \cite{Derman:1979nf}).
	This classification of vacua \cite{Emmanuel-Costa:2016vej} uses the notation R-X-y,
	explained in the text. The vevs of the $\phi_i$ are denoted by $\rho_i$ .}
\label{Table:real}
\begin{center}
\begin{tabular}{|c|c|c|c|}
\hline
\hline
Vacuum  & $\rho_1,\rho_2,\rho_3$ & $w_1,w_2,w_S$ & Comment  \\
\hline
\hline
R-0& $0,0,0$  & $0,0,0$ & Not interesting  \\
\hline
\hline
R-I-1 & $x,x,x$  & $0,0,w_S$ & $\mu _0^2= -\lambda _8 w_S^2$  \\
\hline
R-I-2a & $x,-x,0$ & $w,0,0$ &  $\mu _1^2=- \left(\lambda _1+\lambda _3\right) w_1^2$\\
\hline
R-I-2b & $x,0,-x$ & $w, \sqrt{3} w, 0$ & $\mu _1^2= -\frac{4}{3} \left(\lambda _1+\lambda _3\right) w_2^2$ \\
\hline
R-I-2c & $0,x,-x$ & $w, -\sqrt{3} w, 0$ &$\mu _1^2= -\frac{4}{3}\left(\lambda _1+\lambda _3\right) w_2^2$  \\
\hline
\hline
R-II-1a & $x,x,y$ & $0,w,w_S$ & $\mu _0^2= \frac{1}{2}\lambda _4\frac{ w_2^3}{w_S}
-\frac{1}{2} \lambda_a w_2^2-\lambda _8 w_S^2$, \\
& & & $\mu _1^2= -\left( \lambda _1+ \lambda _3\right) w_2^2+\frac{3}{2} \lambda _4 w_2 w_S-\frac{1}{2} \lambda_a w_S^2$\\
\hline
R-II-1b & $x,y,x$ & $w,-w/\sqrt{3},w_S$ & $\mu _0^2= -4\lambda _4\frac{ w_2^3}{w_S}-2\lambda_a w_2^2-\lambda _8 w_S^2$, \\
& & & $\mu _1^2=-4 \left(\lambda _1+\lambda _3\right) w_2^2 -3 \lambda _4 w_2 w_S-\frac{1}{2} \lambda_a w_S^2$\\
\hline
R-II-1c & $y,x,x$ & $w,w/\sqrt{3},w_S$ &  $\mu _0^2= -4\lambda _4\frac{w_2^3}{w_S}-2\lambda_a w_2^2-\lambda _8 w_S^2$, \\
& & & $\mu _1^2= -4 \left(\lambda _1+\lambda _3\right) w_2^2-3 \lambda _4 w_2 w_S-\frac{1}{2} \lambda_a w_S^2$\\
\hline
R-II-2 & $x,x,-2x$ & $0, w, 0$ &$\mu _1^2= - \left(\lambda _1+\lambda _3\right) w_2^2$, $\lambda_4=0$  \\
\hline
R-II-3 & $x,y,-x-y$ & $w_1,w_2,0$ & $\mu _1^2= -\left(\lambda _1+\lambda _3\right)(w_1^2+w_2^2),\lambda_4=0$ \\
\hline
\hline
R-III & $\rho_1,\rho_2,\rho_3$ & $w_1,w_2,w_S$ & $\mu _0^2= -\frac{1}{2} \lambda_a( w_1^2+ w_2^2)-\lambda _8 w_S^2$,\\
& & & $\mu _1^2= - \left( \lambda _1+ \lambda _3\right)( w_1^2+ w_2^2)-\frac{1}{2} \lambda_a w_S^2$, \\
& & & $\lambda_4=0$ \\
\hline
\hline
\end{tabular}
\end{center}
\end{table}
%%%%%%%%%%%%%%%%%%%%%%%%%%%%%%%%%%%%%%%%%%%%%

%%%%%%%%%%%%%%%%%%%%%%%%%%%%%%%%%%%%%%%%%%%%%
\begin{table}[htb]
\caption{Complex vacua. Notation: $\epsilon=1$ and $-1$ for C-III-d and C-III-e, respectively;
$\xi=\sqrt{-3\sin 2\rho_1/\sin2\rho_2}$,
${\psi=\sqrt{[3+3\cos (\rho_2-2 \rho _1)]/(2\cos\rho_2)}}$. Imposing the vacuum constraints \cite{Emmanuel-Costa:2016vej}, the vacua labelled with an asterisk ($^\ast$) are in fact real.}
\label{Table:complex}
\begin{center}
\begin{tabular}{|c|c|c|}
\hline\hline
& Irreducible Rep.& Defining Rep. \\
\hline
& $w_1,w_2,w_S$ & $\rho_1,\rho_2,\rho_3$  \\
\hline
\hline
C-I-a & $\hat w_1,\pm i\hat w_1,0$ & 
$x, xe^{\pm\frac{2\pi i}{3}}, xe^{\mp\frac{2\pi i}{3}}$ \\
\hline
\hline
C-III-a & $0,\hat w_2e^{i\sigma_2},\hat w_S$ & $y, y, xe^{i\tau}$  \\
\hline
C-III-b & $\pm i\hat w_1,0,\hat w_S$ & $x+iy,x-iy,x$  \\
\hline
C-III-c & $\hat w_1 e^{i\sigma_1},\hat w_2e^{i\sigma_2},0$ 
& $xe^{i\rho}-\frac{y}{2}, -xe^{i\rho}-\frac{y}{2}, y$  \\
\hline
C-III-d,e & $\pm i \hat w_1,\epsilon\hat w_2,\hat{w}_S$ & $xe^{ i\tau},xe^{- i\tau},y$ \\
\hline
C-III-f & $\pm i\hat w_1 ,i\hat w_2,\hat{w}_S$ 
& $re^{i\rho}\pm ix,re^{i\rho}\mp ix,\frac{3}{2}re^{-i\rho}-\frac{1}{2}re^{i\rho}$ \\
\hline
C-III-g & $\pm i\hat w_1,-i\hat w_2,\hat{w}_S$ 
& $re^{-i\rho}\pm ix,re^{-i\rho}\mp ix,\frac{3}{2}re^{i\rho}-\frac{1}{2}re^{-i\rho}$ \\
\hline
C-III-h & $\sqrt{3}\hat w_2 e^{i\sigma_2},\pm\hat w_2 e^{i\sigma_2},\hat{w}_S$ 
& $xe^{i\tau} , y , y$ \\
& & $y, xe^{i\tau},y$ \\
\hline
C-III-i & $\sqrt{\frac{3(1+\tan^2\sigma_1)}{1+9\tan^2\sigma_1}}\hat w_2e^{i\sigma_1},$ 
& $x, ye^{i\tau},ye^{-i\tau}$ \\
& $\pm\hat w_2e^{-i\arctan(3\tan\sigma_1)},\hat w_S$ 
& $ye^{i\tau}, x, ye^{-i\tau}$ \\
\hline
\hline
C-IV-a$^\ast$ & $\hat w_1e^{i\sigma_1},0,\hat w_S$ & $re^{i\rho}+x, -re^{i\rho}+x,x$ \\
\hline
C-IV-b & $\hat w_1,\pm i\hat w_2,\hat w_S$ 
& $re^{i\rho}+x, -re^{-i\rho}+x, -re^{i\rho}+re^{-i\rho}+x$ \\
\hline
C-IV-c & $\sqrt{1+2\cos^2\sigma_2}\hat w_2,$ &  $re^{i\rho}+r\sqrt{3(1+2\cos^2\rho)}+x$, \\
& $\hat w_2e^{i\sigma_2},\hat w_S$ 
& $re^{i\rho}-r\sqrt{3(1+2\cos^2\rho)}+x,-2re^{i\rho}+x$ \\
\hline
C-IV-d$^\ast$ & $\hat w_1e^{i\sigma_1},\pm\hat w_2e^{i\sigma_1},\hat w_S$ & $r_1e^{i\rho}+x, (r_2-r_1)e^{i\rho}+x,-r_2e^{i\rho}+x$ \\
\hline
C-IV-e & $\sqrt{-\frac{\sin 2\sigma_2}{\sin 2\sigma_1}}\hat w_2e^{i\sigma_1},$ & $re^{i\rho_2}+re^{i\rho_1}\xi+x, re^{i\rho_2}-re^{i\rho_1}\xi+x,$  \\
& $\hat w_2e^{i\sigma_2},\hat w_S$ & $-2re^{i\rho_2}+x$ \\
\hline
C-IV-f & $\sqrt{2+\frac{\cos \left(\sigma _1-2 \sigma _2\right)}{\cos\sigma_1}}\hat w_2e^{i\sigma_1},$ & $re^{i\rho_1}+re^{i\rho_2}\psi+x$, \\
& $\hat w_2e^{i\sigma_2},\hat w_S$ &$re^{i\rho_1}-re^{i\rho_2}\psi+x, -2re^{i\rho_1}+x$ \\
\hline
\hline
C-V$^\ast$ & $\hat w_1e^{i\sigma_1},\hat w_2e^{i\sigma_2},\hat w_S$ & $xe^{i\tau_1},ye^{i\tau_2},z$ \\
\hline
\end{tabular}
\end{center}
\end{table}

%%%%%%%%%%%%%%%%%%%%%%%%%%%%%%%%%%%%%%%%%%%%%
\section{Masses in the softly broken C-III-c models}
\setcounter{equation}{0}
\label{app:B}
%%%%%%%%%%%%%%%%%%%%%%%%%%%%%%%%%%%%%%%%%%%%%

Regardless of the softly broken $S_3$ parameters, the charged mass eigenstates are of the same form in every softly broken C-III-c model. Therefore, only the neutral states are analysed here.
Also, since $\hat{w}_S=0$, the masses will only depend on the relative phase,
\begin{equation}
\sigma\equiv\sigma_1-\sigma_2,
\end{equation}
not on $\sigma_1$ and $\sigma_2$ separately.

%%%%%%%%%%%%%%%%%%%%%%%%%%%%%%%%%%%%%%%%%%%%%
\subsection{Models with \boldmath$\lambda_4=0$ and \boldmath$\lambda_2+ \lambda_3 \neq0$}
%%%%%%%%%%%%%%%%%%%%%%%%%%%%%%%%%%%%%%%%%%%%%
These are the models discussed in section~\ref{sect:lam4=0}.
The softly broken parameters consistent with the $\lambda_4=0$ constraint are $\mu_2^2$ and $\nu_{12}^2$. Due to the fact that $\lambda_4=0$ and $\hat{w}_S=0$, there is no mixing between the $S_3$ doublet and singlet at the level of masses. 
These models might provide possible dark matter candidates. Since the $S_3$ singlet possesses zero vacuum value it could be associated with an inert doublet. Moreover, the $\lambda_4=0$ constraint results in a discrete symmetry  $\mathbb{Z}_2:~h_S\to-h_S$, which could stabilise the scalar dark matter sector.

%%%%%%%%%%%%%%%%%%%%%%%%%%%%%%%%%%%%%%%%%%%%%
\subsubsection{Case~2. C-III-c-$\mu_2^2$}
%%%%%%%%%%%%%%%%%%%%%%%%%%%%%%%%%%%%%%%%%%%%%

CP is conserved, and the squared masses are given by:
\begin{subequations}
\begin{align}
m_{H_{(1,2)}}^2 & = \left( \lambda_1 + \lambda_3  \right) v^2 \mp \Delta ,\\
m_A^2 & = 2 \left( \lambda_2 + \lambda_3 \right) v^2 ,\\
m_{S_{(1,2)}}^2 &= \mu_0^2 +  \frac{1}{2} \left(\lambda_5 + \lambda_6 \right)v^2\mp\lambda_7(\hat w_1^2-\hat w_2^2),
\end{align}
\end{subequations}
where
\begin{equation}
\Delta^2 = \left( \lambda_1 + \lambda_3 \right)^2 v^4 - 16 \left( \lambda_1 - \lambda_2 \right)\left( \lambda_2 + \lambda_3 \right) \hat{w}_1^2 \hat{w}_2^2.  
\end{equation}
In the limit $\lambda_2+\lambda_3\to0$, both $m_{H_{1}}^2$ and $m_A^2$ vanish linearly. 

%%%%%%%%%%%%%%%%%%%%%%%%%%%%%%%%%%%%%%%%%%%%%
\subsubsection{Case~3. C-III-c-$\nu_{12}^2$}
%%%%%%%%%%%%%%%%%%%%%%%%%%%%%%%%%%%%%%%%%%%%%

CP is conserved, and the squared masses are given by:
\begin{subequations}
\begin{align}
m_{H_{(1,3)}}^2 &= \left( \lambda_1 + \lambda_3 \mp \Delta \right)v^2 ,\\
m_{H_2}^2 & = 2 \left( \lambda_2 + \lambda_3 \right)v^2,\\
m_{S_{(1,2)}}^2 &= \mu_0^2 + \frac{1}{2} \left( \lambda_5 + \lambda_6 \right)v^2 \mp \lambda_7 \cos \sigma\,v^2,
\end{align}
\end{subequations}
where
\begin{equation}
\Delta^2  =\left( \lambda_1 - \lambda_2 \right)^2 + \left( \lambda_2 + \lambda_3 \right)^2 
+2 \left( \lambda_1 - \lambda_2 \right)\left( \lambda_2 + \lambda_3 \right) \cos (2\sigma).
\end{equation}
The above terminology is determined by $m_{H_1}^2 < m_{H_{2}}^2<m_{H_{3}}^2$, valid for $\mu_1^2<0$.
In the limit $\lambda_2+\lambda_3\to0$, both $m_{H_{1}}^2$ and $m_{H_2}^2$ vanish linearly.

%%%%%%%%%%%%%%%%%%%%%%%%%%%%%%%%%%%%%%%%%%%%%
\subsubsection{Case~4. C-III-c-$\mu_2^2$-$\nu_{12}^2$}
%%%%%%%%%%%%%%%%%%%%%%%%%%%%%%%%%%%%%%%%%%%%%

CP is conserved, illustrating that the conditions (\ref{Eq:first_case}) are not {\it necessary}. This is explained in appendix~\ref{app:C}.
The squared masses are given by:
\begin{subequations}
\begin{align}
m_{H_{(1,2)}}^2 &= \left( \lambda_1 + \lambda_3  \right)v^2 \mp \Delta_H,\\
m_{H_3}^2 & = 2\left( \lambda_2 + \lambda_3 \right)v^2,\\
m_{S_{(1,2)}}^2 &= \mu_0^2 + \frac{1}{2} \left( \lambda_5 + \lambda_6\right) v^2\mp \lambda_7 \Delta_S,
\end{align}
\end{subequations}
where
\begin{align}
\Delta_H^2 &= \left( \lambda_1 + \lambda_3 \right)^2 v^4  
- 16 \left( \lambda_1 - \lambda_2 \right)\left( \lambda_2 + \lambda_3 \right)\sin^2 \sigma \hat{w}_1^2 \hat{w}_2^2,\\
\Delta_S^2 &= v^4-4\sin^2\sigma\,\hat w_1^2\hat w_2^2.
\end{align}
The above terminology is determined by $m_{H_1}^2 < m_{H_{2}}^2<m_{H_{3}}^2$, valid for $\mu_1^2<0$.
In the limit $\lambda_2+\lambda_3\to0$, both $m_{H_{1}}^2$ and $m_{H_3}^2$ vanish linearly.

%%%%%%%%%%%%%%%%%%%%%%%%%%%%%%%%%%%%%%%%%%%%%
\subsection{Models with \boldmath$\lambda_4 \neq 0$ and \boldmath$\lambda_2+ \lambda_3=0$}
%%%%%%%%%%%%%%%%%%%%%%%%%%%%%%%%%%%%%%%%%%%%%
These are the models discussed in section~\ref{sect:lam2+lam3=0}. They actually contain states of negative squared mass, but are described here as limits to be avoided in any discussion of realistic versions of Cases~9, 10 and 11.

The softly broken parameters consistent with the $\lambda_3 + \lambda_2=0$ constraint are $\nu_{01}^2$ and $\nu_{02}^2$. Due to the fact that $\lambda_4 \neq 0$, there is  mixing between the $S_3$ doublet and singlet. 

%%%%%%%%%%%%%%%%%%%%%%%%%%%%%%%%%%%%%%%%%%%%%
\subsubsection{Case~6. C-III-c-$\nu_{02}^2$}
%%%%%%%%%%%%%%%%%%%%%%%%%%%%%%%%%%%%%%%%%%%%%
Because of the mixing between the doublet and singlet sectors, the neutral mass-squared matrix is now $6\times6$. Transforming to the Higgs basis and removing the would-be Goldstone boson, it is reduced to a $5\times5$ matrix spanned by the fields
\begin{equation} \label{Eq:HB-fields}
\{\eta_1^\text{HB},\eta_2^\text{HB},\eta_3^\text{HB},\chi_2^\text{HB},\chi_3^\text{HB}\}.
\end{equation}
The $5\times5$ mass-squared matrix is block-diagonal, a $3\times3$ ``$\eta$'' block and a $2\times2$ ``$\chi$'' block, see Eq.~\eqref{Eq:hi_hS}, reflecting the fact that CP is conserved.
The ``$\eta$'' block has the following properties:
\begin{align}
\det \left( \mathcal{M}_\eta^2 \right) & = 8 (\lambda_2 - \lambda_1) \lambda_4^2 \hat{w}_1^2 \hat{w}_2^4,
\label{Eq:Case6-det-eta}\\
\Tr \left( \mathcal{M}_\eta^2\right) & = \mu_0^2 + 2(\lambda_1 - \lambda_2)v^2  
+ \half(\lambda_5+ \lambda_6) v^2 +\lambda_7(\hat w_2^2-\hat w_1^2).
\end{align}
For the product of the three squared masses to be positive, we must have $\lambda_2>\lambda_1$.
For the ``$\chi$'' block we may solve explicitly for the squared masses in terms of a square root:
\begin{align}
m_{A_{(1,2)}}^2 = \frac{1}{4} \left( 2 \mu_0^2 + \lambda_a \hat{w}_1^2 + \lambda_b \hat{w}_2^2\mp \Delta \right),
\end{align}
where
\begin{equation}
\Delta^2 = 16 \lambda_4^2 \hat{w}_1^2 v^2 + \left( 2 \mu_0^2 + \lambda_a \hat{w}_1^2 + \lambda_b \hat{w}_2^2 \right)^2.
\end{equation}
We note that the product of these two masses is negative,
\begin{equation}
m_{A_1}^2 m_{A_2}^2=-\lambda_4^2\hat w_1^2 v^2 <0,
\end{equation}
so Case~6 must be abandoned. Actually, with $\lambda_1-\lambda_2$ positive (as follows from $\mu_1^2$ being negative), also the determinant (\ref{Eq:Case6-det-eta}) is negative, signalling the fact that not one, but {\it two} masses squared are negative.
%%%%%%%%%%%%%%%%%%%%%%%%%%%%%%%%%%%%%%%%%%%%%
\subsubsection{Case~7. C-III-c-$\nu_{01}^2$}
%%%%%%%%%%%%%%%%%%%%%%%%%%%%%%%%%%%%%%%%%%%%%
In this case, the $5\times5$ matrix spanned by the fields (\ref{Eq:HB-fields})
has the following structure
\begin{equation} \label{Eq:5by5}
\mathcal{M}^2 = \begin{pmatrix}
\times & 0 & \times & 0 &\times \\
0 & 0 & \times & 0 & 0 \\
\times & \times &\times &  \times & \times\\
0 & 0 & \times & 0 & \times\\
\times & 0 & \times & \times & \times
\end{pmatrix}.
\end{equation}
While individual elements (denoted ``$\times$'') also depend on $\mu_0^2$, $\lambda_5$, $\lambda_6$ and $\lambda_7$, the product of all 5 masses squared is very simple,
\begin{equation} \label{Eq:case7-det}
M_1^2M_2^2M_3^2M_4^2M_5^2
=\det \left(\mathcal{M}^2\right) = \frac{1}{2} \left( \lambda_1 - \lambda_2 \right) \lambda_4^4 \sin^4 \sigma\, v^{10}.
\end{equation}
However, the sum of all masses squared depends on these additional parameters,
\begin{equation}
\Tr \left( \mathcal{M}^2 \right) = 2\mu_0^2 +  2(\lambda_1 - \lambda_2)v^2 + (\lambda_5 + \lambda_6)v^2.
\end{equation}
A necessary condition for positive squared masses is $\lambda_1>\lambda_2.$

One might be tempted to conclude from Eq.~(\ref{Eq:case7-det}) that {\it four} of the squared masses vanish as $\lambda_4\to0$. This is not necessarily the case, as illustrated by the following example. Let a toy model have the two squared masses:
\begin{equation}
m_{a,b}^2=\left[\lambda_0\mp\sqrt{\lambda_0^2-\lambda_4^2}\right]v^2.
\end{equation}
While the product is given by $m_{a}^2m_{a}^2=\lambda_4^2\,v^4$, the individual masses squared are for $\lambda_4/\lambda_0\to0$ given as
\begin{equation}
m_a^2\simeq\frac{\lambda_4^2}{2\lambda_0}\,v^2, \quad 
m_b^2\simeq 2\lambda_0\,v^2.
\end{equation}

As mentioned above, in Case~7 and for $\lambda_1>\lambda_2$, the overall determinant is positive, allowing for an even number of negative squared masses. A numerical exploration shows that two of them are negative, like for Case~6.
%%%%%%%%%%%%%%%%%%%%%%%%%%%%%%%%%%%%%%%%%%%%%
\subsubsection{Case~8. C-III-c-$\nu_{01}^2$-$\nu_{02}^2$}
%%%%%%%%%%%%%%%%%%%%%%%%%%%%%%%%%%%%%%%%%%%%%

The Higgs basis rotation and reduction to a $5\times5$ matrix yields the following structure
\begin{equation}
\mathcal{M}^2 = \begin{pmatrix}
\times & 0 & \times & 0 &\times \\
0 & 0 & \times & 0 & \times\\
\times & \times &\times &  \times & \times\\
0 & 0 & \times & 0 & \times\\
\times & \times & \times & \times & \times
\end{pmatrix},
\end{equation}
with
\begin{align}
\det \left( \mathcal{M}^2 \right) & = 8 \left( \lambda_1 - \lambda_2 \right) \lambda_4^4 \sin^4 \sigma \hat{w}_1^4 \hat{w}_2^4 v^2,\\
\Tr \left( \mathcal{M}^2 \right) & = 2\mu_0^2 + \left( 2\lambda_1 - 2\lambda_2 + \lambda_5 + \lambda_6 \right)v^2 .
\end{align}
For positive squared masses we must have $\lambda_1>\lambda_2$.

The mass eigenstates are all mixtures of the gauge fields of Eq.~(\ref{Eq:HB-fields}) and CP is violated.

The overall determinant is positive, allowing for an even number of negative squared masses. A numerical exploration shows that two of them are negative, like for Case~6.

%%%%%%%%%%%%%%%%%%%%%%%%%%%%%%%%%%%%%%%%%%%%%
\subsection{Models with \boldmath$\lambda_4 \neq 0$ and \boldmath$\lambda_2+ \lambda_3 \neq 0$}
%%%%%%%%%%%%%%%%%%%%%%%%%%%%%%%%%%%%%%%%%%%%%
These are the models discussed in section~\ref{sect:lam2+lam3neq0_and_lam4neq0}.
In these models there is mixing between the $S_3$ doublet and singlet.

%%%%%%%%%%%%%%%%%%%%%%%%%%%%%%%%%%%%%%%%%%%%%
\subsubsection{Case~9. C-III-c-$\mu_2^2$-$\nu_{02}^2$}
%%%%%%%%%%%%%%%%%%%%%%%%%%%%%%%%%%%%%%%%%%%%%

The $5\times5$ neutral mass-squared matrix is block-diagonal, a $3\times3$ ``$\eta$'' block and a $2\times2$ ``$\chi$'' block, reflecting the fact that CP is conserved. The following properties can be extracted:
\begin{align}
\det \left( \mathcal{M}^2_\eta \right) & = 8\left( \lambda_1 - \lambda_2 \right)\left[ \left( \lambda_2 + \lambda_3 \right) 
\left( 2\mu_0^2 + \lambda_b \hat{w}_1^2 + \lambda_a \hat{w}_2^2 \right) - \lambda_4^2 \hat{w}_2^2\right] 
\hat{w}_1^2\hat{w}_2^2,\\
\Tr \left( \mathcal{M}^2_\eta \right) & = \mu_0^2 + 2\left(\lambda_1 + \lambda_3 \right) v^2 
+ \half(\lambda_b \hat{w}_1^2 + \lambda_a \hat{w}_2^2).
\end{align}
The masses squared of the ``$\chi$'' sector are given by:
\begin{equation}
m_{A_{(1,2)}}^2 = \frac{1}{4}\left[ 2\mu_0^2 + 4\left( \lambda_2 + \lambda_3 + \frac{1}{4} \lambda_a \right)\hat{w}_1^2 + 4\left( \lambda_2 + \lambda_3 + \frac{1}{4} \lambda_b \right)\hat{w}_2^2 \mp \Delta\right],
\end{equation}
where
\begin{equation}
\begin{split}
\Delta^2 &= \left[ 2 \mu_0^2 + 4\left( \lambda_2 + \lambda_3 + \frac{1}{4} \lambda_a \right)\hat{w}_1^2 + 4\left( \lambda_2 + \lambda_3 + \frac{1}{4} \lambda_b \right)\hat{w}_2^2 \right]^2 \\
&\hspace{10pt}- 16 \left[ \left( 2\mu_0^2 + \lambda_a \hat{w}_1^2 + \lambda_b \hat{w}_2^2 \right)\left( \lambda_2 + \lambda_3 \right) - \lambda_4^2 \hat{w}_1^2 \right]v^2.
\end{split}
\end{equation}
For the ``$\chi$'' sector we have
\begin{equation}
\det \left( \mathcal{M}^2_\chi \right)
=\left[\left( \lambda_2 + \lambda_3 \right) \left( 2\mu_0^2 + \lambda_a \hat{w}_1^2 + \lambda_b \hat{w}_2^2 \right) - \lambda_4^2 \hat{w}_1^2\right]v^2.
\end{equation}
Necessary conditions to have all squared masses positive is then
\begin{align}
\left( \lambda_2 + \lambda_3 \right) 
\left( 2\mu_0^2 + \lambda_b \hat{w}_1^2 + \lambda_a \hat{w}_2^2 \right) - \lambda_4^2 \hat{w}_2^2&>0, \\
\left( \lambda_2 + \lambda_3 \right) \left( 2\mu_0^2 + \lambda_a \hat{w}_1^2 + \lambda_b \hat{w}_2^2 \right) - \lambda_4^2 \hat{w}_1^2&>0.
\end{align}

%%%%%%%%%%%%%%%%%%%%%%%%%%%%%%%%%%%%%%%%%%%%%
\subsubsection{Case~10. C-III-c-$\nu_{12}^2$-$\nu_{01}^2$}
%%%%%%%%%%%%%%%%%%%%%%%%%%%%%%%%%%%%%%%%%%%%%

In this case, the $5\times5$ neutral mass-squared matrix takes the form:
\begin{equation}
\mathcal{M}^2_\mathrm{0} = \begin{pmatrix}
\times & 0 & \times & \times &\times \\
0 & \times & \times & 0 & 0\\
\times & \times &\times &  \times & \times\\
\times & 0 & \times & \times & \times\\
\times & 0 & \times & \times & \times
\end{pmatrix},
\end{equation}
with
\begin{equation}
\det\left( \mathcal{M}^2 \right)=\sin^2\sigma(\lambda_1-\lambda_2)
[A_4(\mu_0^2)^2+A_2\mu_0^2+A_0]v^6,
\end{equation}
where
\begin{subequations}
\begin{align}
A_4&=8(\lambda_2+\lambda_3)^2, \\
A_2&=4(\lambda_2+\lambda_3)
[2(\lambda_2+\lambda_3)(\lambda_5+\lambda_6)-\lambda_4^2] v^2, \\
A_0&=\half\{
[2(\lambda_2+\lambda_3)(\lambda_5+\lambda_6)-\lambda_4^2]^2
-\cos^2\sigma[4(\lambda_2+\lambda_3)\lambda_7-\lambda_4^2]^2\}v^4 \\
&=\half \sin^2\sigma\lambda_4^4 v^4
-2(\lambda_2+\lambda_3)\lambda_4^2
(\lambda_5+\lambda_6-2\lambda_7 \cos^2\sigma)v^4 \nonumber \\
&\quad+2(\lambda_2+\lambda_3)^2[(\lambda_5+\lambda_6)^2-4\lambda_7^2\cos^2\sigma]v^4,
\end{align}
\end{subequations}
and
\begin{equation}
\Tr \left( \mathcal{M}^2 \right) = 2\mu_0^2 
+ \left(2 \lambda _1+2 \lambda _2+4 \lambda _3+\lambda _5+\lambda _6\right) v^2. \label{Eq:trace-M_sq}
\end{equation}

The mass eigenstates are a mixture of all five gauge fields, and CP is violated.

%%%%%%%%%%%%%%%%%%%%%%%%%%%%%%%%%%%%%%%%%%%%%
\subsubsection{Case~11. C-III-c-$\mu_2^2$-$\nu_{12}^2$-$\nu_{01}^2$-$\nu_{02}^2$}
%%%%%%%%%%%%%%%%%%%%%%%%%%%%%%%%%%%%%%%%%%%%%

In this case, all elements of the $5\times5$ neutral mass-squared matrix are non-zero, CP is violated and 
the determinant is rather complex,
\begin{equation}
\det\left( \mathcal{M}^2 \right)=8\sin^2\sigma(\lambda_1-\lambda_2)
[A_4(\mu_0^2)^2+A_2\mu_0^2+A_0]\hat w_1^2\hat w_2^2 v^2,
\end{equation}
with
\begin{subequations}
\begin{align}
A_4&=4(\lambda_2+\lambda_3)^2, \\
A_2&=2(\lambda_2+\lambda_3)
[2(\lambda_2+\lambda_3)(\lambda_5+\lambda_6)-\lambda_4^2] v^2, \\
A_0&=\sin^2\sigma\lambda_4^4\hat w_1^2\hat w_2^2
-(\lambda_2+\lambda_3)\lambda_4^2
[(\lambda_5+\lambda_6)v^4-2(v^4-4\sin^2\sigma\hat w_1^2\hat w_2^2)\lambda_7] \nonumber \\
&+(\lambda_2+\lambda_3)^2[(\lambda_5+\lambda_6)^2v^4-4(v^4-4\sin^2\sigma\hat w_1^2\hat w_2^2)\lambda_7^2] \\
&=(\lambda_2 + \lambda_3)[(\lambda_2 + \lambda_3)\lambda_a - \lambda_4^2]\lambda_b v^4 + \sin^2\sigma[4(\lambda_2 + \lambda_3)\lambda_7 - \lambda_4^2]^2\hat{w}_1^2\hat{w}_2^2,
\end{align}
\end{subequations}
whereas the trace has the familiar value,
given by Eq.~(\ref{Eq:trace-M_sq}).

%%%%%%%%%%%%%%%%%%%%%%%%%%%%%%%%%%%%%%%%%%%%%
\subsection{Sums of masses squared}
%%%%%%%%%%%%%%%%%%%%%%%%%%%%%%%%%%%%%%%%%%%%%
The contributions to the trace of the mass-squared matrix in the neutral sector, i.e., the sum of all squared masses of the neutral scalars can be expressed in terms of $\mu_0^2$, $\mu_1^2$, $\lambda_2+\lambda_3$, $\lambda_5+\lambda_6$, and $\lambda_7$ as
\begin{equation}
\Tr\left( \mathcal{M}^2 \right)=
\#\mu_0^2+\#\mu_1^2+\#\half(\lambda_2+\lambda_3)v^2+\#\half(\lambda_5+\lambda_6)v^2
+\#\lambda_7v^2,
\end{equation}
where we use the minimisation condition $\mu_1^2=-(\lambda_1-\lambda_2)v^2$ and trade $\lambda_3$ for $\mu_1^2$ and $(\lambda_2+\lambda_3)$.

We summarise in table~\ref{Table:trM_sq} the coefficients denoted ``$\#$'' above. Where CP is conserved, contributions to the CP-even and CP-odd parts are given separately, as $x+y$. 
The singlet sector contributes to the $\mu_0^2$, $(\lambda_5+\lambda_6)$ and $\lambda_7$ terms (the latter cancel among CP-even and odd terms), whereas the doublet sector contributes to the $\mu_1^2$ and $(\lambda_2+\lambda_3)$ terms.

%%%%%%%%%%%%%%%%%%%%%%%%%%%%%%%%%%%%%%%%%%%%%
\begin{table}[htb]
\caption{Contributions to sums of masses squared.}
\label{Table:trM_sq}
\begin{center}
\begin{tabular}{|c|c|c|c|c|c|c|c|c|c|}
\hline\hline
Case &CPC & $\mu_0^2$ & $\mu_1^2$ & $\half(\lambda_2+\lambda_3)$ 
& $\half(\lambda_5+\lambda_6)$& $\lambda_7$\\
\hline
\hline
1 & $\checkmark$ & 1+1 & $-2+0$ & $0$ & 1+1 & $\pm a\mp a$\\
\hline
2 &$\checkmark$  & 1+1 & $-2+0$ & 4+4 & 1+1 & $\pm b\mp b$\\
\hline
3 & $\checkmark$ & 1+1 & $-2+0$ & 4+4 & 1+1 & $\pm a\mp a$\\
\hline
4 & $\checkmark$ & 1+1 & $-2+0$ & 4+4 & 1+1 & $\pm c\mp c$\\
\hline
5 &$\checkmark$  & 1+1 & $-2+0$ & 0 & 1+1 & $-b+b$ \\
\hline
6 & $\checkmark$  & 1+1 & $-2+0$ & 0 & 1+1 & $-b+b$ \\
\hline
7 & $-$ & 2 & $-2$ & 0 & 2 & 0 \\
\hline
8 & $-$ & 2 & $-2$ & 0 & 2 & 0 \\
\hline
9 & $\checkmark$ & 1+1 & $-2+0$ & 4+4 & 1+1 & $-b+b$ \\
\hline
10 & $-$ & 2 & $-2$ & 8 & 2 & 0 \\
\hline
11 & $-$ & 2 & $-2$ & 8 & 2 & 0 \\
\hline
\hline
\end{tabular}
\end{center}
$a=\cos\sigma\, v^2$, $b=(\hat w_1^2-\hat w_2^2)$,
$c=\Delta_S$.
\end{table}
%%%%%%%%%%%%%%%%%%%%%%%%%%%%%%%%%%%%%%%%%%%%%

Expressed in these terms, the structure is remarkably simple and regular.
%%%%%%%%%%%%%%%%%%%%%%%%%%%%%%%%%%%%%%%%%%%%%
\section{CP conservation in Case~4}
\setcounter{equation}{0}
\label{app:C}
%%%%%%%%%%%%%%%%%%%%%%%%%%%%%%%%%%%%%%%%%%%%%

For Case 4, when the two soft-breaking terms $\mu_2^2$ and $\nu_{12}^2$ are present, the vacuum configuration $(\hat w_1e^{i\sigma_1},\hat w_2e^{i\sigma_2},0)$ does not result in a CP-violating model. This can be shown explicitly by constructing a basis transformation that results in both a real potential and a real vacuum, thereby eliminating the possibility of having spontaneous CP violation. We start by simultaneously rephasing all three doublets to get a vacuum configuration of the form $(\hat w_1e^{i\sigma}, \hat w_2,0)$, leaving all parameters of the potential 
\begin{align}
	V&=\mu_0^2 h_S^\dagger h_S +\mu_1^2(h_1^\dagger h_1 + h_2^\dagger h_2)+\mu_2^2(h_1^\dagger h_1 - h_2^\dagger h_2)+\frac{1}{2}\nu_{12}^2(h_2^\dagger h_1 + h_1^\dagger h_2) \\
	&+\lambda_1(h_1^\dagger h_1 + h_2^\dagger h_2)^2 
	+\lambda_2(h_1^\dagger h_2 - h_2^\dagger h_1)^2
	+\lambda_3[(h_1^\dagger h_1 - h_2^\dagger h_2)^2+(h_1^\dagger h_2 + h_2^\dagger h_1)^2]
	\nonumber \\
	&+\lambda_5(h_S^\dagger h_S)(h_1^\dagger h_1 + h_2^\dagger h_2)
	+\lambda_6[(h_S^\dagger h_1)(h_1^\dagger h_S)+(h_S^\dagger h_2)(h_2^\dagger h_S)] \nonumber \\
	&+\lambda_7[(h_S^\dagger h_1)(h_S^\dagger h_1) + (h_S^\dagger h_2)(h_S^\dagger h_2) +\text{h.c.}]
	+\lambda_8(h_S^\dagger h_S)^2.
	\end{align}
unchanged.
Consider now the change of basis given by
\begin{eqnarray}
\left(
\begin{array}{c}\bar{h}_1\\ \bar{h}_2 \\ \bar{h}_S
\end{array}\right)
&=&
e^{i\psi}\left(
\begin{array}{ccc}\cos\theta & e^{-i\xi}\sin\theta & 0\\ -e^{i\chi}\sin\theta & e^{i(\chi-\xi)}\cos\theta & 0\\
	0 & 0 & e^{i\tau}
\end{array}\right)
\left(
\begin{array}{c}h_1\\ h_2 \\ h_S
\end{array}\right),
\end{eqnarray}
with 
\begin{eqnarray}
\theta&=&\arctan\left(\frac{\hat w_2}{\hat w_1}\right),\\
\chi&=&-\arctan\left(\frac{v^2}{(\hat w_2^2-\hat w_1^2)\tan\sigma}\right),\\
\xi&=&-\sigma,\\
\psi&=&-\sigma,\\
\tau&=&\frac{\pi}{4}+\frac{\sigma}{2}-\frac{1}{2}\arctan\left(\frac{v^2}{(\hat w_2^2- \hat w_1^2)\tan\sigma}\right).
\end{eqnarray}	
This takes us to the Higgs basis, with the real vacuum configuration $(v,0,0)$ along with a transformed potential of the form
\begin{eqnarray}
	\bar{V} &=&\bar{\gamma}_0 (\bar{h}_S^\dagger \bar{h}_S) +\bar{\gamma}_1 (\bar{h}_1^\dagger \bar{h}_1)  +\bar{\gamma}_2  (\bar{h}_2^\dagger \bar{h}_2)
	+\bar{\gamma}_3  \left[(\bar{h}_1^\dagger \bar{h}_2)+(\bar{h}_2^\dagger \bar{h}_1)\right]
	\nonumber\\
	&& + \frac{\bar{\Lambda}_1}{2}(\bar{h}_1^\dagger \bar{h}_1)^2
	+ \frac{\bar{\Lambda}_2}{2}(\bar{h}_2^\dagger \bar{h}_2)^2
+	\bar{\Lambda}_3(\bar{h}_1^\dagger \bar{h}_1)(\bar{h}_2^\dagger \bar{h}_2)
	+	\bar{\Lambda}_4(\bar{h}_1^\dagger \bar{h}_2)(\bar{h}_2^\dagger \bar{h}_1)\nonumber\\
	&&+\frac{\bar{\Lambda}_5}{2}\left[(\bar{h}_2^\dagger \bar{h}_1)^2+(\bar{h}_1^\dagger \bar{h}_2)^2
	\right]
	+\bar{\Lambda}_6\left[(\bar{h}_1^\dagger \bar{h}_1)(\bar{h}_1^\dagger \bar{h}_2)+(\bar{h}_1^\dagger \bar{h}_1)(\bar{h}_2^\dagger \bar{h}_1)
	\right]\nonumber\\
	&&+\bar{\Lambda}_7\left[(\bar{h}_2^\dagger \bar{h}_2)(\bar{h}_1^\dagger \bar{h}_2)+(\bar{h}_2^\dagger \bar{h}_2)(\bar{h}_2^\dagger \bar{h}_1)\right]
	+\bar{\Lambda}_8(\bar{h}_S^\dagger \bar{h}_S)
	\left[(\bar{h}_1^\dagger \bar{h}_1)+(\bar{h}_2^\dagger \bar{h}_2)\right]\nonumber\\
	&&+\bar{\Lambda}_9\left[(\bar{h}_S^\dagger \bar{h}_1)(\bar{h}_1^\dagger \bar{h}_S)+(\bar{h}_S^\dagger \bar{h}_2)(\bar{h}_2^\dagger \bar{h}_S)\right]
	+\frac{\bar{\Lambda}_{10}}{2}\left[(\bar{h}_S^\dagger \bar{h}_1)^2+(\bar{h}_1^\dagger \bar{h}_S)^2\right]\nonumber\\
	&&+\frac{\bar{\Lambda}_{11}}{2}\left[(\bar{h}_S^\dagger \bar{h}_2)^2+(\bar{h}_2^\dagger \bar{h}_S)^2
	\right]
	+\frac{\bar{\Lambda}_{12}}{2}\left[(\bar{h}_S^\dagger \bar{h}_1)(\bar{h}_S^\dagger \bar{h}_2)+(\bar{h}_1^\dagger \bar{h}_S)(\bar{h}_2^\dagger \bar{h}_S)
	\right]\nonumber\\
	&&+\bar{\Lambda}_{13}(\bar{h}_S^\dagger \bar{h}_S)^2
\end{eqnarray}
where all the $\bar\gamma_i$ and $\bar\Lambda_i$ become real when imposing Eq.~(\ref{Eq:minimisation-soft}), thus implying a CP conserving model.
%%%%%%%%%%%%%%%%%%%%%%%%%%%%%%%%%%%%%%%%%%%%%
\section{Continuous symmetries of the potential}
\setcounter{equation}{0}
\label{app:D}
%%%%%%%%%%%%%%%%%%%%%%%%%%%%%%%%%%%%%%%%%%%%%
Symmetries of multi-Higgs-doublet models are only manifest in particular bases. Ref.~\cite{deMedeirosVarzielas:2019rrp} provides a very useful prescription to 
identify the symmetries present in specific implementations of three-Higgs-doublet models when written in  bases where these symmetries are not manifest. It is well known that a $\mathbb{Z}_2$ symmetry acting on one Higgs doublet in models with two Higgs doublets may manifest itself as a symmetry for the interchange of the two doublets. Here we show how a continuous O(2) symmetry of an $S_3$-symmetric potential corresponding to the C-III-c vacuum can explicitly appear as a different continuous symmetry.

Let us consider the C-III-c vacuum configuration $(\hat w_1e^{i\sigma_1},\hat w_2e^{i\sigma_2},0)$, in the framework of the $S_3$-symmetric 3HDM potential without soft breaking terms.
The stationary-point equations require $\lambda_4=0$ and $\lambda_3=-\lambda_2$, so
the quartic part of the potential simplifies to
\begin{eqnarray}
V_4&=&
\lambda_1(h_1^\dagger h_1 + h_2^\dagger h_2)^2 
-\lambda_2[(h_1^\dagger h_1 - h_2^\dagger h_2)^2+4(h_1^\dagger h_2)(h_2^\dagger h_1)]
%\nonumber \\
%&& 
+\lambda_5(h_S^\dagger h_S)(h_1^\dagger h_1 + h_2^\dagger h_2) \nonumber \\
&&+\lambda_6[(h_S^\dagger h_1)(h_1^\dagger h_S)+(h_S^\dagger h_2)(h_2^\dagger h_S)] 
+\lambda_7[(h_S^\dagger h_1)^2 + (h_S^\dagger h_2)^2+(h_1^\dagger h_S)^2 + (h_2^\dagger h_S)^2]
\nonumber \\
&&+\lambda_8(h_S^\dagger h_S)^2.
\end{eqnarray}
Using the basis-independent checks given in \cite{deMedeirosVarzielas:2019rrp}, we find that our potential satisfies the constraints for both O(2) and U(1)$_1$ symmetries. The O(2) symmetry is manifest in the basis in which we start out because the potential is symmetric now under the change of basis given by
\begin{equation} 
\left( \begin{array}{c}
h_1\\
h_2\\
h_S \\
\end{array}  \right) \to O
\left( \begin{array}{c}
h_1\\
h_2\\
h_S \\
\end{array}  \right),
\end{equation}
where
\begin{equation}\label{O2}O\in\left\{
\left( \begin{array}{ccc} 
\cos\alpha & -\sin\alpha & 0 \\ 
\sin\alpha & \cos\alpha & 0 \\
0 & 0 & 1 \\
\end{array} \right) 
,\\
\left( \begin{array}{ccc} 
\cos\alpha & \sin\alpha & 0 \\ 
\sin\alpha & -\cos\alpha & 0 \\
0 & 0 & 1 \\
\end{array} \right)\right\}.
\end{equation}
The U(1)$_1$ symmetry does not manifest itself in this basis, so it must be a hidden symmetry which manifests itself in another basis. Let us therefore change into another basis using
\begin{equation} 
\left( \begin{array}{c}
h_1\\
h_2\\
h_S \\
\end{array}  \right) = B
\left( \begin{array}{c}
\phi_1 \\
\phi_2 \\
\phi_3 \\
\end{array}  \right),
\end{equation}
where
\begin{equation} \label{Eq:transform-B}
B = \left( \begin{array}{ccc} 
\frac{1}{\sqrt{2}} & \frac{1}{\sqrt{2}} & 0 \\ 
-\frac{i}{\sqrt{2}} & \frac{i}{\sqrt{2}} & 0 \\
0 & 0 & 1 \\
\end{array} \right).
\end{equation}
The transformed potential in the new basis is given by
\begin{eqnarray}
V_2&=&\mu_0^2 \phi_3^\dagger \phi_3 +\mu_1^2(\phi_1^\dagger \phi_1 + \phi_2^\dagger \phi_2), \\
V_4&=&
\lambda_1(\phi_1^\dagger \phi_1 + \phi_2^\dagger \phi_2)^2 
-\lambda_2[(\phi_1^\dagger \phi_1 - \phi_2^\dagger \phi_2)^2+4(\phi_1^\dagger \phi_2)(\phi_2^\dagger \phi_1)]
%\nonumber \\
%&& 
+\lambda_5(\phi_3^\dagger \phi_3)(\phi_1^\dagger \phi_1 + \phi_2^\dagger \phi_2) \nonumber \\
&&+\lambda_6[(\phi_3^\dagger \phi_1)(\phi_1^\dagger \phi_3)+(\phi_3^\dagger \phi_2)(\phi_2^\dagger \phi_3)] 
+2\lambda_7[(\phi_1^\dagger \phi_3)(\phi_2^\dagger \phi_3) + (\phi_3^\dagger \phi_1)(\phi_3^\dagger \phi_2)] + \lambda_8(\phi_3^\dagger \phi_3)^2.
\end{eqnarray}
Applying the transformation (\ref{Eq:transform-B}) to the vacuum (\ref{Eq:vac-sigmahalf}), it is seen to become real.

In this new basis, the potential is manifestly symmetric under the U(1)$_1$ transformation given by
\begin{equation} 
\left( \begin{array}{c}
\phi_1 \\
\phi_2 \\
\phi_3 \\
\end{array}  \right) \to \left( \begin{array}{ccc} 
e^{i\alpha} & 0 & 0 \\ 
0 & e^{-i\alpha} & 0 \\
0 & 0 & 1 \\
\end{array} \right)
\left( \begin{array}{c}
\phi_1 \\
\phi_2 \\
\phi_3 \\
\end{array}  \right).
\end{equation}
This U(1)$_1$ symmetry is not an additional continuous symmetry, it is just the way the original O(2) symmetry manifests itself after the change of basis. 
If the potential possesses a symmetry in the original basis represented by a matrix $S$, then after changing to a new basis, the potential will possess a symmetry represented by the matrix $B^\dagger SB$. Applying this to the matrices in (\ref{O2}), we find that the O(2) matrices are transformed into the following two matrices in the new basis
\begin{equation}
\left\{\left( \begin{array}{ccc} 
e^{i\alpha} & 0 & 0 \\ 
0 & e^{-i\alpha} & 0 \\
0 & 0 & 1 \\
\end{array} \right) 
,\\
\left( \begin{array}{ccc} 
0 & e^{i\alpha} & 0 \\ 
e^{-i\alpha} & 0 & 0 \\
0 & 0 & 1 \\
\end{array} \right)\right\},
\end{equation}
thus showing that this U(1)$_1$ symmetry is just the original O(2) symmetry expressed in the new basis.
  
%%%%%%%%%%%%%%%%%%%%%%%%%%%%%%%%%%%%%%%%%%%%%


\begin{thebibliography}{99}

%\cite{Higgs:1964ia}
\bibitem{Higgs:1964ia}
  P.~W.~Higgs,
  {\it Broken symmetries, massless particles and gauge fields,}
  Phys.\ Lett.\  {\bf 12} (1964) 132.
  %%CITATION = doi:10.1016/0031-9163(64)91136-9;%%
  
 %\cite{Higgs:1964pj}
\bibitem{Higgs:1964pj}
  P.~W.~Higgs,
  {\it Broken Symmetries and the Masses of Gauge Bosons,}
  Phys.\ Rev.\ Lett.\  {\bf 13} (1964) 508.
  %%CITATION = doi:10.1103/PhysRevLett.13.508;%%
 
 %\cite{Englert:1964et}
\bibitem{Englert:1964et}
  F.~Englert and R.~Brout,
  {\it Broken Symmetry and the Mass of Gauge Vector Mesons,}
  Phys.\ Rev.\ Lett.\  {\bf 13} (1964) 321.
  %%CITATION = doi:10.1103/PhysRevLett.13.321;%%

%\cite{Guralnik:1964eu}
\bibitem{Guralnik:1964eu}
  G.~S.~Guralnik, C.~R.~Hagen and T.~W.~B.~Kibble,
   {\it Global Conservation Laws and Massless Particles,}
  Phys.\ Rev.\ Lett.\  {\bf 13} (1964) 585.
  %doi:10.1103/PhysRevLett.13.585
  %%CITATION = doi:10.1103/PhysRevLett.13.585;%%

 %\cite{Aad:2012tfa}
\bibitem{Aad:2012tfa}
  G.~Aad {\it et al.} [ATLAS Collaboration],
{\it Observation of a new particle in the search for the Standard Model Higgs boson with the ATLAS detector at the LHC,}
  Phys.\ Lett.\ B {\bf 716} (2012) 1
  [arXiv:1207.7214 [hep-ex]].
  %%CITATION = doi:10.1016/j.physletb.2012.08.020;%%

%\cite{Chatrchyan:2012xdj}
\bibitem{Chatrchyan:2012xdj}
  S.~Chatrchyan {\it et al.} [CMS Collaboration],
 {\it Observation of a new boson at a mass of 125 GeV with the CMS experiment at the LHC,}
  Phys.\ Lett.\ B {\bf 716} (2012) 30
  [arXiv:1207.7235 [hep-ex]].
  %%CITATION = doi:10.1016/j.physletb.2012.08.021;%%
  %%CITATION = FRPHA,80,1;%%

%\cite{Gunion:1989we}
\bibitem{Gunion:1989we}
  J.F.~Gunion, H.E.~Haber, G.L.~Kane and S.~Dawson,
  \textit{The Higgs Hunter's Guide}
  (Westview Press, Boulder, CO, 2000).

%\cite{Branco:2011iw}
\bibitem{Branco:2011iw}
  G.C.~Branco, P.M.~Ferreira, L.~Lavoura, M.N.~Rebelo, M.~Sher and J.P.~Silva,
 {\it Theory and phenomenology of two-Higgs-doublet models,}
  Phys.\ Rept.\  {\bf 516} (2012) 1
   [arXiv:1106.0034 [hep-ph]].
  %%CITATION = doi:10.1016/j.physrep.2012.02.002;%%

%\cite{Ivanov:2017dad}
\bibitem{Ivanov:2017dad}
  I.~P.~Ivanov,
  {\it Building and testing models with extended Higgs sectors,}
  Prog.\ Part.\ Nucl.\ Phys.\  {\bf 95} (2017) 160
   [arXiv:1702.03776 [hep-ph]].
  %%CITATION = doi:10.1016/j.ppnp.2017.03.001;%%
 
 %\cite{Olaussen:2010aq}
\bibitem{Olaussen:2010aq}
  K.~Olaussen, P.~Osland and M.~A.~Solberg,
  {\it Symmetry and Mass Degeneration in Multi-Higgs-Doublet Models,}
  JHEP {\bf 1107} (2011) 020
   [arXiv:1007.1424 [hep-ph]].
  %%CITATION = doi:10.1007/JHEP07(2011)020;%%
  
%\cite{Lee:1973iz}
\bibitem{Lee:1973iz}
  T.~D.~Lee,
 {\it A Theory of Spontaneous T Violation,}
  Phys.\ Rev.\ D {\bf 8} (1973) 1226.
  %%CITATION = PHRVA,D8,1226;%%

%\cite{Nambu:1960tm}
\bibitem{Nambu:1960tm}
  Y.~Nambu,
   {\it Quasiparticles and Gauge Invariance in the Theory of Superconductivity,}
  Phys.\ Rev.\  {\bf 117} (1960) 648.
  % doi:10.1103/PhysRev.117.648
  %%CITATION = doi:10.1103/PhysRev.117.648;%%

%\cite{Goldstone:1961eq}
\bibitem{Goldstone:1961eq}
  J.~Goldstone,
   {\it Field Theories with Superconductor Solutions,}
  Nuovo Cim.\  {\bf 19} (1961) 154.
  %doi:10.1007/BF02812722
  %%CITATION = doi:10.1007/BF02812722;%%

%\cite{Goldstone:1962es}
\bibitem{Goldstone:1962es}
  J.~Goldstone, A.~Salam and S.~Weinberg,
  {\it Broken Symmetries,}
  Phys.\ Rev.\  {\bf 127} (1962) 965.
  % doi:10.1103/PhysRev.127.965
  %%CITATION = doi:10.1103/PhysRev.127.965;%%


%\cite{Tanabashi:2018oca}
\bibitem{Tanabashi:2018oca}
  M.~Tanabashi {\it et al.} [Particle Data Group],
  {\it Review of Particle Physics,}
  Phys.\ Rev.\ D {\bf 98} (2018) no.3,  030001.
   %%CITATION = doi:10.1103/PhysRevD.98.030001;%%

%\cite{Emmanuel-Costa:2016vej}
\bibitem{Emmanuel-Costa:2016vej}
  D.~Emmanuel-Costa, O.~M.~Ogreid, P.~Osland and M.~N.~Rebelo,
  {\it Spontaneous symmetry breaking in the $S_3$-symmetric scalar sector,}
  JHEP {\bf 1602} (2016) 154
   Erratum: [JHEP {\bf 1608} (2016) 169]
  [arXiv:1601.04654 [hep-ph]].
  %%CITATION = doi:10.1007/JHEP08(2016)169, 10.1007/JHEP02(2016)154;%%

%\cite{Pakvasa:1977in}
\bibitem{Pakvasa:1977in}
  S.~Pakvasa and H.~Sugawara,
 {\it Discrete Symmetry and Cabibbo Angle,}
  Phys.\ Lett.\  {\bf 73B} (1978) 61.
  %%CITATION = doi:10.1016/0370-2693(78)90172-7;%%

%\cite{Derman:1978rx}
\bibitem{Derman:1978rx}
  E.~Derman,
{\it Flavor Unification, $\tau$ Decay and $b$ Decay Within the Six Quark Six Lepton {Weinberg-Salam} Model,}
  Phys.\ Rev.\ D {\bf 19} (1979) 317.
  %%CITATION = PHRVA,D19,317;%%

%\cite{Derman:1979nf}
\bibitem{Derman:1979nf}
  E.~Derman and H.~S.~Tsao,
  {\it SU(2)$\times$U(1)$\times$S($n$) Flavor Dynamics and a Bound on the Number of Flavors,}
  Phys.\ Rev.\ D {\bf 20} (1979) 1207.
  %%CITATION = PHRVA,D20,1207;%%

\bibitem{Kuncinas}
A. Kuncinas, 
{\it Properties of S3-Symmetric Three-Higgs-Doublet Models,}
Master thesis, University of Bergen, 2019;
http://bora.uib.no/handle/1956/20467.

%\cite{deMedeirosVarzielas:2019rrp}
\bibitem{deMedeirosVarzielas:2019rrp}
  I.~de Medeiros Varzielas and I.~P.~Ivanov,
  {\it Recognizing symmetries in a 3HDM in a basis-independent way,}
  Phys.\ Rev.\ D {\bf 100} (2019) no.1,  015008
  % doi:10.1103/PhysRevD.100.015008
  [arXiv:1903.11110 [hep-ph]].
  %%CITATION = doi:10.1103/PhysRevD.100.015008;%%

%\cite{Darvishi:2019dbh}
\bibitem{Darvishi:2019dbh}
  N.~Darvishi and A.~Pilaftsis,
 {\it Classifying Accidental Symmetries in Multi-Higgs Doublet Models,}
  arXiv:1912.00887 [hep-ph].
  %%CITATION = ARXIV:1912.00887;%%
  
%\cite{Ishimori:2012zz}
\bibitem{Ishimori:2012zz}
  H.~Ishimori, T.~Kobayashi, H.~Ohki, H.~Okada, Y.~Shimizu and M.~Tanimoto,
  {\it An introduction to non-Abelian discrete symmetries for particle physicists,}
  Lect.\ Notes Phys.\  {\bf 858} (2012) 1.
  %%CITATION = doi:10.1007/978-3-642-30805-5;%%

%\cite{Kubo:2004ps}
\bibitem{Kubo:2004ps}
  J.~Kubo, H.~Okada and F.~Sakamaki,
  {\it Higgs potential in minimal S(3) invariant extension of the standard model,}
  Phys.\ Rev.\ D {\bf 70} (2004) 036007
  [hep-ph/0402089].
  %%CITATION = HEP-PH/0402089;%%

%\cite{Teshima:2012cg}
\bibitem{Teshima:2012cg}
  T.~Teshima,
  {\it Higgs potential in $S_3$ invariant model for quark/lepton mass and mixing,}
  Phys.\ Rev.\ D {\bf 85} (2012) 105013
  [arXiv:1202.4528 [hep-ph]].
  %%CITATION = ARXIV:1202.4528;%%

%\cite{Das:2014fea}
\bibitem{Das:2014fea}
  D.~Das and U.~K.~Dey,
  {\it Analysis of an extended scalar sector with $S_3$ symmetry,}
  Phys.\ Rev.\ D {\bf 89} (2014) 095025
 [Phys.\ Rev.\ D {\bf 91} (2015) 3,  039905]
  [arXiv:1404.2491 [hep-ph]].
  %%CITATION = ARXIV:1404.2491;%%

%\cite{Ferreira:2004yd}
\bibitem{Ferreira:2004yd}
P.~Ferreira, R.~Santos and A.~Barroso,
%``Stability of the tree-level vacuum in two Higgs doublet models against charge or CP spontaneous violation,''
Phys.\ Lett.\ B \textbf{603} (2004), 219-229
doi:10.1016/j.physletb.2004.10.022
[arXiv:hep-ph/0406231 [hep-ph]].
%146 citations counted in INSPIRE as of 04 Apr 2020

%\cite{Barroso:2005sm}
\bibitem{Barroso:2005sm}
  A.~Barroso, P.~M.~Ferreira and R.~Santos,
  %``Charge and CP symmetry breaking in two Higgs doublet models,''
  Phys.\ Lett.\ B {\bf 632} (2006) 684
  doi:10.1016/j.physletb.2005.11.031
  [hep-ph/0507224].
  %%CITATION = doi:10.1016/j.physletb.2005.11.031;%%
  %73 citations counted in INSPIRE as of 06 Apr 2020
  
%\cite{Ivanov:2010wz}
\bibitem{Ivanov:2010wz}
I.~Ivanov,
%``Properties of the general NHDM. II. Higgs potential and its symmetries,''
JHEP \textbf{07} (2010), 020
doi:10.1007/JHEP07(2010)020
[arXiv:1004.1802 [hep-th]].
%31 citations counted in INSPIRE as of 04 Apr 2020

%\cite{Ogreid:2017alh}
\bibitem{Ogreid:2017alh}
  O.~M.~Ogreid, P.~Osland and M.~N.~Rebelo,
  {\it A Simple Method to detect spontaneous CP Violation in multi-Higgs models,}
  JHEP {\bf 1708} (2017) 005
  [arXiv:1701.04768 [hep-ph]].
  %%CITATION = doi:10.1007/JHEP08(2017)005;%%
 
%\cite{Branco:1983tn}
\bibitem{Branco:1983tn} 
  G.~C.~Branco, J.~M.~Gerard and W.~Grimus,
  {\it Geometrical T Violation,}
  Phys.\ Lett.\ B {\bf 136}, 383 (1984).
  %%CITATION = PHLTA,B136,383;%%

%\cite{Donoghue:1978cj}
\bibitem{Donoghue:1978cj}
  J.~F.~Donoghue and L.~F.~Li,
  {\it Properties of Charged Higgs Bosons,}
  Phys.\ Rev.\ D {\bf 19} (1979) 945.
  %%CITATION = doi:10.1103/PhysRevD.19.945;%%

%\cite{Georgi:1978ri}
\bibitem{Georgi:1978ri}
  H.~Georgi and D.~V.~Nanopoulos,
  {\it Suppression of Flavor Changing Effects From Neutral Spinless Meson Exchange in Gauge Theories,}
  Phys.\ Lett.\ B {\bf 82} (1979) 95.
  %%CITATION = doi:10.1016/0370-2693(79)90433-7;%%

%\cite{Branco:1985aq}
\bibitem{Branco:1985aq}
  G.~C.~Branco and M.~N.~Rebelo,
  {\it The Higgs Mass in a Model With Two Scalar Doublets and Spontaneous {CP} Violation,}
  Phys.\ Lett.\  {\bf 160B} (1985) 117.
  %%CITATION = doi:10.1016/0370-2693(85)91476-5;%%

%\cite{Haber:2018iwr}
\bibitem{Haber:2018iwr}
  H.~E.~Haber, O.~M.~Ogreid, P.~Osland and M.~N.~Rebelo,
  {\it Symmetries and Mass Degeneracies in the Scalar Sector,}
  JHEP {\bf 1901} (2019) 042
    [arXiv:1808.08629 [hep-ph]].
  %%CITATION = doi:10.1007/JHEP01(2019)042;%%

%\cite{Chakrabarty:2015kmt}
\bibitem{Chakrabarty:2015kmt}
  N.~Chakrabarty,
  {\it High-scale validity of a model with Three-Higgs-doublets,}
  Phys.\ Rev.\ D {\bf 93} (2016) no.7,  075025
  % doi:10.1103/PhysRevD.93.075025
  [arXiv:1511.08137 [hep-ph]].
  %%CITATION = doi:10.1103/PhysRevD.93.075025;%%


%\cite{Das:2015sca}
\bibitem{Das:2015sca}
  D.~Das, U.~K.~Dey and P.~B.~Pal,
  {\it $S_3$ symmetry and the quark mixing matrix,}
  Phys.\ Lett.\ B {\bf 753} (2016) 315
  % doi:10.1016/j.physletb.2015.12.038
  [arXiv:1507.06509 [hep-ph]].
  %%CITATION = doi:10.1016/j.physletb.2015.12.038;%%

%\cite{Gomez-Izquierdo:2018jrx}
\bibitem{Gomez-Izquierdo:2018jrx}
  J.~C.~G\' omez-Izquierdo and M.~Mondrag\' on,
  {\it B–L Model with $\mathbf{S}_{3}$ symmetry: Nearest Neighbor Interaction 
Textures and Broken $\mu\leftrightarrow\tau$ Symmetry,}
  Eur.\ Phys.\ J.\ C {\bf 79} (2019) no.3,  285
  % doi:10.1140/epjc/s10052-019-6785-5
  [arXiv:1804.08746 [hep-ph]].
  %%CITATION = doi:10.1140/epjc/s10052-019-6785-5;%%

%\cite{Chakrabarty:2019tsm}
\bibitem{Chakrabarty:2019tsm}
  N.~Chakrabarty and I.~Chakraborty,
  {\it Flavour-alignment in an $S_3$-symmetric Higgs sector and its RG-behaviour,}
  arXiv:1903.09388 [hep-ph].
  %%CITATION = ARXIV:1903.09388;%%

%\cite{Das:2019yad}
\bibitem{Das:2019yad}
  D.~Das and I.~Saha,
   {\it Alignment limit in three Higgs-doublet models,}
  Phys.\ Rev.\ D {\bf 100} (2019) no.3,  035021
  % doi:10.1103/PhysRevD.100.035021
  [arXiv:1904.03970 [hep-ph]].
  %%CITATION = doi:10.1103/PhysRevD.100.035021;%%

  
%\cite{Machado:2012ed}
\bibitem{Machado:2012ed}
  A.~C.~B.~Machado and V.~Pleitez,
  {\it A model with two inert scalar doublets,}
  Annals Phys.\  {\bf 364} (2016) 53
  %doi:10.1016/j.aop.2015.10.017
  [arXiv:1205.0995 [hep-ph]].
  %%CITATION = doi:10.1016/j.aop.2015.10.017;%%

  
%\cite{Fortes:2014dca}
\bibitem{Fortes:2014dca}
  E.~C.~F.~S.~Fortes, A.~C.~B.~Machado, J.~Montaño and V.~Pleitez,
  {\it Scalar dark matter candidates in a two inert Higgs doublet model,}
  J.\ Phys.\ G {\bf 42} (2015) no.10,  105003
  %doi:10.1088/0954-3899/42/10/105003
  [arXiv:1407.4749 [hep-ph]].
  %%CITATION = doi:10.1088/0954-3899/42/10/105003;%%


%\cite{Espinoza:2018itz}
\bibitem{Espinoza:2018itz}
  C.~Espinoza, E.~A.~Garc\' es, M.~Mondrag\' on and H.~Reyes-Gonz\' alez,
   {\it The $S3$ Symmetric Model with a Dark Scalar,}
  Phys.\ Lett.\ B {\bf 788} (2019) 185
  % doi:10.1016/j.physletb.2018.11.028
  [arXiv:1804.01879 [hep-ph]].
  %%CITATION = doi:10.1016/j.physletb.2018.11.028;%%


%\cite{Mishra:2019keq}
\bibitem{Mishra:2019keq}
  S.~Mishra,
  {\it Majorana dark matter and neutrino mass with $S_3$ symmetry,}
  arXiv:1911.02255 [hep-ph]. 
  %%CITATION = ARXIV:1911.02255;%%

%\cite{Gerard:1982mm}
\bibitem{Gerard:1982mm}
  J.~M.~Gerard,
  {\it Fermion Mass Spectrum in $SU(2)_L \times U(1)$,}
  Z.\ Phys.\ C {\bf 18} (1983) 145.
  % doi:10.1007/BF01572477
  %%CITATION = doi:10.1007/BF01572477;%%

%\cite{Barradas-Guevara:2015rea}
\bibitem{Barradas-Guevara:2015rea}
  E.~Barradas-Guevara, O.~F\' elix-Beltr\' an and E.~Rodr\'\i guez-J\' auregui,
   {\it CP breaking in $S(3)$ flavoured Higgs model,}
  arXiv:1507.05180 [hep-ph].
  %%CITATION = ARXIV:1507.05180;%%



\end{thebibliography}
\end{document}